\documentclass[twocolumn]{jpsj3}

\usepackage{color}
\usepackage{bm}

\title{%
Andreev Reflection in a Bilayer Graphene Junction:\\
Role of Spatial Variation of the Charge Neutrality Point
}

\author{%
Yositake Takane$^1$, Katsuhide Yarimizu$^2$, and Akinobu Kanda$^2$
}

\inst{%
$^1$Department of Quantum Matter,
Graduate School of Advanced Sciences of Matter,\\
Hiroshima University, Higashihiroshima, Hiroshima 739-8530, Japan\\
$^2$Division of Physics and TIMS, Faculty of Pure and Applied Sciences,\\
University of Tsukuba, Tsukuba 305-8571, Japan
}

\recdate{ \hspace{50mm} }

\abst{%
A graphene sheet partially covered with a bulk superconductor serves as
a normal conductor--superconductor (NS) junction, in which electron transport
is mainly governed by Andreev reflection (AR).
As excess carriers induced over the covered region penetrate into
the uncovered region over a screening length, the charge neutrality point
(CNP) in the uncovered region shifts only near the NS interface.
We theoretically study the electron transport in a bilayer graphene junction
taking account of such spatial variation of the CNP
in the electron-doped case.
When the Fermi level is close to the CNP away from the NS interface,
the AR takes place in a specular manner owing to the diffraction of
a reflected hole occurring at a $pn$ junction,
which is naturally formed in the uncovered region.
It is shown that the differential conductance shows
an unusual asymmetric behavior as a function of bias voltage
under the influence of the $pn$ junction.
It is also shown that, if the Fermi level is located below the CNP,
the $pn$ junction gives rise to quasi-bound states near the NS interface,
leading to the appearance of
resonant peaks in the differential conductance.
}

\begin{document}
\sloppy
\maketitle

\section{Introduction}

A normal conductor--superconductor (NS) junction of graphene
can be fabricated by partially covering a graphene sheet
with a bulk superconductor,~\cite{heersche,miao,sato,du,jeong,shalom,borzenets}
where the covered and uncovered regions
respectively correspond to S and N electrodes.
It has been suggested that an NS junction of graphene shows
an anomalous feature in the quasiparticle scattering process
at an NS interface, which cannot be observed in an ordinary NS junction
composed by connecting a normal metal with a superconductor.

Let us consider the case where an electron is incident to the NS interface
from the N side and its energy $E$ measured from the Fermi level $E_{\rm F}$
is in the subgap region of
$|E| < \Delta_{0}$ with $\Delta_{0}$ being the pair potential.
Owing to the presence of the energy gap, an incident electron is
reflected back as either an electron or a hole.
The ordinary reflection process from an electron to an electron is called
normal reflection (NR).
The off-diagonal scattering process from an electron to a hole
is called Andreev reflection (AR),~\cite{andreev}
which dominates the transport properties of quasiparticles.~\cite{BTK,van-wees}
In an ordinary NS junction, a reflected hole traces back the path of
an incident electron (i.e., retroreflection).
This feature stems from the fact that both an incident electron and
a reflected hole are in the conduction band.
Contrastingly, in a graphene NS junction, an incident electron
in the conduction band can be reflected back as a hole in the valence band
if $E_{\rm F}$ is close to the charge neutrality point (CNP),
which is set to be zero hereafter, such that $\Delta_{0} > E > |E_{\rm F}|$.
In this case, an incident electron is reflected as a hole
in a specular manner.~\cite{beenakker}
This phenomenon called specular AR has attracted significant
attention from both experimentalists and theorists.
However, it has not been clearly observed in an experiment.
A central reason for this is that its experimental detection necessitates
an extremely clean sample with almost no potential fluctuations
to satisfy $\Delta_{0} > E > |E_{\rm F}|$.
This condition requires that potential fluctuations
$\delta E_{\rm F}$ in space must be much smaller than $\Delta_{0}$.

Recently, a high-quality graphene sheet has been realized by supporting it
with hexagonal boron nitride.~\cite{dean}
Indeed, a sample with $\delta E_{\rm F} \sim 5$ meV
has been reported in Ref.~\citen{xue}.
Using a proximity system of such a high-quality bilayer graphene sheet,
Efetov \textit{et al.}~\cite{efetov1} have experimentally studied
the differential conductance in a graphene NS junction.
Here, bilayer graphene rather than monolayer graphene is utilized as
$\delta E_{\rm F}$ is smaller in the former than in the latter.
They observed suppression of the differential conductance
in the subgap region when $E_{\rm F}$ is tuned close to the CNP,
and claimed that this suppression is a characteristic feature of specular AR.
However, in order to solidly confirm their conclusion, we need to resolve
problems that remain to be clarified.

\begin{figure}[bp]
\begin{center}
\includegraphics[height=4.5cm]{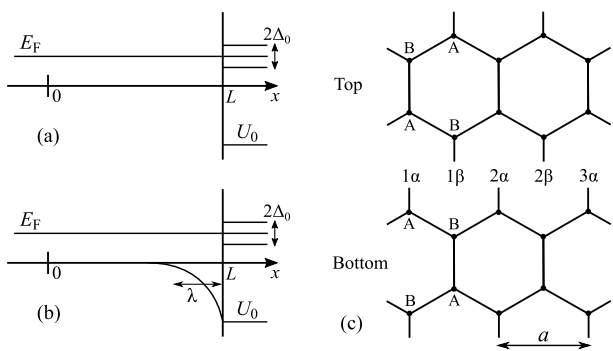}
\end{center}
\caption{Potential profile in a bilayer graphene junction
in the electron-doped case
where the screening length is (a) zero and (b) finite.
(c) Hexagonal lattice system with lattice constant $a$ for bilayer graphene
consisting of top and bottom layers, where an A sublattice site
on the top layer is located exactly on a B sublattice site in the bottom layer.
}
\end{figure}

Previous theoretical analyses on an NS junction of
bilayer graphene~\cite{efetov1,efetov2,ludwig} assumed that the CNP varies
in a stepwise manner across the interface between the covered and uncovered
regions (i.e., piecewise position-independent) [see Fig.~1(a)].
However, the CNP near the NS interface should be position-dependent,
particularly in the uncovered region,
reflecting the inhomogeneity of the carrier concentration [see Fig.~1(b)].
Indeed, excess carriers are induced over the covered region of $x > L$
from a superconductor and then inevitably penetrate into the uncovered region
of $x < L$ up to a screening length $\lambda$
from the NS interface.~\cite{mueller}
Thus, we expect that, in the electron (hole)-doped case,
the CNP in the uncovered region significantly shifts
in the downward (upward) direction near the NS interface over $\lambda$
and gradually approaches the asymptotic value, being set equal to zero,
with increasing distance from the NS interface.
It is important to examine the effect of such spatial variation of the CNP.

In this paper, we numerically calculate the differential conductance of
a graphene junction in the presence of spatial variation of the CNP
by using a tight-binding model.
Considering the status of experimental efforts in this field,
we focus on NS junctions composed of bilayer graphene.
In particular, our attention is directed to the electron-doped case
where $E_{\rm F}$ in the covered region is located far enough above the CNP
while $E_{\rm F}$ in the uncovered region is close to the CNP
in the asymptotic region of $x \le 0$ far away from the NS interface
[see Fig.~1(b)].
In this setup, the AR takes place as ordinary retroreflection
since an electron and a hole in the conduction band
are involved in the AR process near the NS interface.
Despite this, the AR can be viewed as specular reflection
if observed in the asymptotic region.
This is attributed to the diffraction of a reflected hole
due to a smooth $pn$ junction,~\cite{cheianov1,cheianov2}
which is naturally formed in the uncovered region.
It is shown that the differential conductance shows
an unusual asymmetric behavior as a function of bias voltage
under the influence of the $pn$ junction.
It is also shown that, if $E_{\rm F}$ is located below the CNP,
resonant peaks appear in the differential conductance,
reflecting the presence of quasi-bound states created by the $pn$ junction
together with the NS interface.

In the next section, we introduce a tight-binding model
for the bilayer graphene junction and briefly describe a numerical method
to calculate the differential conductance of the junction.
In Sect.~3, we present the numerical results of the differential conductance
for several sets of parameters and explain their characteristics.
The last section is devoted to a summary and discussion.
Differences between this study and that of Ref.~\citen{efetov2}
are briefly discussed with an emphasis on the role of $\lambda$.
We set $\hbar = 1$ throughout this paper.

\section{Tight-Binding Model}

We consider a bilayer graphene sheet of width $W$
that is infinitely long in the $x$-direction, 
where the covered region of $x > L$ corresponds to the S electrode
and the uncovered region of $x < L$ corresponds to the N electrode.
For simplicity, we assume that the potential $V(x)$
determining the location of the CNP is given by
\begin{align}
  V(x)
  = \left\{ \begin{array}{cc}
               -U_{0} \, e^{(x-L)/\lambda} & {\rm if} \quad x \le L , \\
               -U_{0} & {\rm if} \quad x > L , \\
            \end{array}
    \right.
\end{align}
with $L \gg \lambda$.
The potential vanishes in the region of $x \le 0$,
which is regarded as an asymptotic region.
The pair potential induced in a bilayer graphene sheet is
simply assumed as~\cite{comment1}
\begin{align}
    \label{eq:Delta-def}
  \Delta(x)
  = \left\{ \begin{array}{cc}
               0 & {\rm if} \quad x \le L , \\
               \Delta_{0} & {\rm if} \quad x > L . \\
            \end{array}
    \right.
\end{align}
With this setup, we consider AR and NR for an electron with energy $E$
measured from the Fermi energy $E_{\rm F}$
incident from the asymptotic region of $x \le 0$.

Let us introduce a pair of vertically coupled hexagonal lattices,
as shown in Fig.~1(c), on which a bilayer graphene is implemented within
a tight-binding approximation.~\cite{wallace,slonczewski,mccann}
A unit cell consists of two different columns containing four lattice sites:
a pair of A and B sublattice sites in the top and bottom layers.
We designate them as the $\alpha$ and $\beta$ columns; thus, each column in
this lattice system is specified by this index together with the number $j$
of the unit cell (i.e., $j \alpha$ or $j \beta$).
The coordinate $x$ of the $j \alpha$ ($j \beta$) column is
given by $x = aj$ [$x = a(j+1/2)$] with $a$ being the lattice constant.
Now, let us introduce the vector of wave functions in the $j \zeta$ column
($\zeta = \alpha$ or $\beta$) as
\begin{align}
    \label{eq:vec-e+h}
 \hat{\mib c}_{j}^{\zeta}
  = \left[ \begin{array}{c}
               \mib{e}_{j}^{\zeta} \\
               \mib{h}_{j}^{\zeta} \\
            \end{array}
    \right] ,
\end{align}
where $\mib{e}_{j}^{\zeta}$ and $\mib{h}_{j}^{\zeta}$ are respectively
the electron and hole wave functions given by
\begin{align}
 \mib{e}_{j}^{\alpha}
  = \left[ \begin{array}{c}
               e_{j{\rm A}}^{\alpha} \\
               e_{j{\rm B}}^{\alpha} \\
               e_{j{\rm A^{'}}}^{\alpha} \\
               e_{j{\rm B^{'}}}^{\alpha} \\
            \end{array}
    \right] ,
 \hspace{6mm}
 \mib{h}_{j}^{\alpha}
  = \left[ \begin{array}{c}
               h_{j{\rm A}}^{\alpha} \\
               h_{j{\rm B}}^{\alpha} \\
               h_{j{\rm A^{'}}}^{\alpha} \\
               h_{j{\rm B^{'}}}^{\alpha} \\
            \end{array}
    \right] .
\end{align}
and
\begin{align}
 \mib{e}_{j}^{\beta}
  = \left[ \begin{array}{c}
               e_{j{\rm B}}^{\beta} \\
               e_{j{\rm A}}^{\beta} \\
               e_{j{\rm B^{'}}}^{\beta} \\
               e_{j{\rm A^{'}}}^{\beta} \\
            \end{array}
    \right] ,
 \hspace{6mm}
 \mib{h}_{j}^{\beta}
  = \left[ \begin{array}{c}
               h_{j{\rm B}}^{\beta} \\
               h_{j{\rm A}}^{\beta} \\
               h_{j{\rm B^{'}}}^{\beta} \\
               h_{j{\rm A^{'}}}^{\beta} \\
            \end{array}
    \right] .
\end{align}
Here, $\rm A$ and $\rm B$ respectively represent the A and B sublattice sites
in the top layer while $\rm A^{'}$ and $\rm B^{'}$ respectively represent
those in the bottom layer.
Using a simple tight-binding approximation, we describe electrons in bilayer
graphene in terms of transfer integrals $\gamma_{0}$ and $\gamma_{1}$,
which respectively represent the in-plane hopping of an electron
between nearest-neighbor sites in each layer and the vertical hopping of
an electron between nearest-neighbor sites in the top and bottom layers.
This approximation ignores the effect of trigonal warping, which is not
expected to significantly affect AR processes.
Under the periodic boundary condition in the $y$-direction,
quasiparticle wave functions in the corresponding direction are
described by a wave number $q$.
In terms of the vector of wave functions defined in Eq.~(\ref{eq:vec-e+h}),
the Bogoliubov--de Gennes equation for a quasiparticle with energy $E$
is expressed as follows:
\begin{align}
    \label{eq:BdG-eq}
   E \hat{\mib c}_{j}^{\alpha}
 & = \hat{\mib H}_{j}^{\alpha}\hat{\mib c}_{j}^{\alpha}
     + \hat{\mib V}\hat{\mib c}_{j}^{\beta}
     + \hat{\mib V}\hat{\mib c}_{j-1}^{\beta} ,
            \\
   E \hat{\mib c}_{j}^{\beta}
 & = \hat{\mib H}_{j}^{\beta}\hat{\mib c}_{j}^{\beta}
     + \hat{\mib V}\hat{\mib c}_{j+1}^{\alpha}
     + \hat{\mib V}\hat{\mib c}_{j}^{\alpha} ,
\end{align}
where
\begin{align}
   \hat{\mib H}_{j}^{\zeta}
 & = \left[ \begin{array}{cc}
               \mib{H}_{j}^{\zeta} & \mib{\Delta} \\
               \mib{\Delta} & -\mib{H}_{j}^{\zeta} \\
            \end{array}
     \right] ,
            \\
   \hat{\mib V}
 & = \left[ \begin{array}{cc}
               \mib{V} & \mib{0} \\
               \mib{0} & -\mib{V} \\
            \end{array}
     \right] .
\end{align}
Here, the submatrices are given by
\begin{align}
   \mib{H}_{j}^{\alpha}
 & = \left[ \begin{array}{cccc}
              -E_{{\rm F}j} & -\gamma_{0} & 0 & \gamma_{1} \\
              -\gamma_{0} & -E_{{\rm F}j} & 0 & 0   \\
               0 & 0 & -E_{{\rm F}j} & -\eta_{q}\gamma_{0} \\
               \gamma_{1} & 0 & -\eta_{q}^{*}\gamma_{0} & -E_{{\rm F}j} \\
            \end{array}
     \right] ,
                 \\
   \mib{H}_{j}^{\beta}
 & = \left[ \begin{array}{cccc}
              -E_{{\rm F}j} & -\eta_{q}^{*}\gamma_{0} & 0 & 0 \\
              -\eta_{q}\gamma_{0} & -E_{{\rm F}j} & \gamma_{1} & 0   \\
               0 & \gamma_{1} & -E_{{\rm F}j} & -\gamma_{0} \\
               0 & 0 & -\gamma_{0} & -E_{{\rm F}j} \\
            \end{array}
     \right]
\end{align}
with
\begin{align}
  E_{{\rm F}j} & = E_{\rm F} - V(x_{j}) ,
             \\
  \eta_{q} & = e^{i\sqrt{3}qa} ,
\end{align}
and
\begin{align}
   \mib{\Delta}
 & = \left[ \begin{array}{cccc}
               \Delta_{0} & 0 & 0 & 0 \\
               0 & \Delta_{0} & 0 & 0 \\
               0 & 0 & \Delta_{0} & 0 \\
               0 & 0 & 0 & \Delta_{0} \\
            \end{array}
     \right] ,
            \\
   \mib{V}
 & = \left[ \begin{array}{cccc}
               -\gamma_{0} & 0 & 0 & 0 \\
               0 & -\gamma_{0} & 0 & 0 \\
               0 & 0 & -\gamma_{0} & 0 \\
               0 & 0 & 0 & -\gamma_{0} \\
            \end{array}
     \right] .
\end{align}

Within the model Hamiltonian described above, we consider
the scattering problem for an electron incident to the NS interface
at $x = L$ from the asymptotic region of $x \le 0$.
In the low-energy region of $\gamma_{1} > |E_{\rm F}+E|$,
the number of conducting channels is $2$, representing the degeneracy of
two energy valleys, in the asymptotic region regardless of $q$.
Using a recursive Green's function technique, we can numerically determine
the $2 \times 2$ AR amplitude $r_{\rm A}(E_{\rm F}+E)$
and the $2 \times 2$ NR amplitude $r_{\rm N}(E_{\rm F}+E)$.
The differential conductance as a function of bias voltage $V$ is given
by the Blonder--Tinkham--Klapwijk formula:~\cite{BTK,takane} 
\begin{align}
       \label{eq:BTK-formula}
  G_{\rm NS}(E_{\rm F}+eV)
  = \frac{2e^{2}}{\pi}\frac{W k_{\rm F}(E_{\rm F}+eV)}{\pi}
    T_{\rm NS}(E_{\rm F}+eV)
\end{align}
with
\begin{align}
   T_{\rm NS}(E_{\rm F}+eV)
 & = \frac{1}{2}
     \int_{0}^{\frac{\pi}{2}} d\phi \cos\phi
        \nonumber \\
 &   \hspace{-6mm} \times
     {\rm tr}\left.
             \left\{ 1_{2 \times 2} - r_{\rm N}^{\dagger}r_{\rm N}
                     + r_{\rm A}^{\dagger}r_{\rm A}
             \right\}\right|_{E_{\rm F}+eV} ,
\end{align}
where $1_{2 \times 2}$ is the $2 \times 2$ unit matrix.
The Fermi wave number $k_{\rm F}$ in the asymptotic region is obtained as
\begin{align}
 k_{\rm F}(E)a
   = \frac{2}{\sqrt{3}}
    \arccos \left(1-\frac{(2|E|+\gamma_{1})^{2}
                     -\gamma_{1}^{2}}{8\gamma_{0}^{2}}\right) ,
\end{align}
and $\phi$ is the angle of incidence defined by $\phi = \arccos(q/k_{\rm F})$.
Here, $W k_{\rm F}/\pi$ in Eq.~(\ref{eq:BTK-formula}) is the total number of
conducting channels in each valley and $T_{\rm NS}$ represents
the angle-averaged dimensionless conductance,
which reaches its maximum value of $2$
when the AR probability becomes $1$ (i.e., perfect AR) regardless of $\phi$.
The factor of $1/2$ in the expression for $T_{\rm NS}$ is included to pick up
the contribution from a single valley.
Although the physical meaning of $T_{\rm NS}$ is transparent,
it cannot fully capture the $V$ dependence of $G_{\rm NS}$.
Thus, we mainly consider $g_{\rm NS}$ defined by
\begin{align}
   g_{\rm NS}(E_{\rm F}+eV) = k_{\rm F}(E_{\rm F}+eV)a
                              T_{\rm NS}(E_{\rm F}+eV) ,
\end{align}
in terms of which the differential conductance is expressed as
\begin{align}
  G_{\rm NS}(E_{\rm F}+eV)
   = \frac{2e^{2}W}{\pi^{2}a} g_{\rm NS}(E_{\rm F}+eV) .
\end{align}
This indicates that $g_{\rm NS}$ can be regarded as
the dimensionless conductance per unit width.
For reference, we also introduce $g_{\rm N}(E_{\rm F}+eV)$ and
$T_{\rm N}(E_{\rm F}+eV)$, each of which represents
the corresponding dimensionless conductance
in the normal state (i.e., $\Delta_{0} = 0$).

\section{Numerical Results}

\begin{figure}[bpt]
\begin{center}
\includegraphics[height=4.0cm]{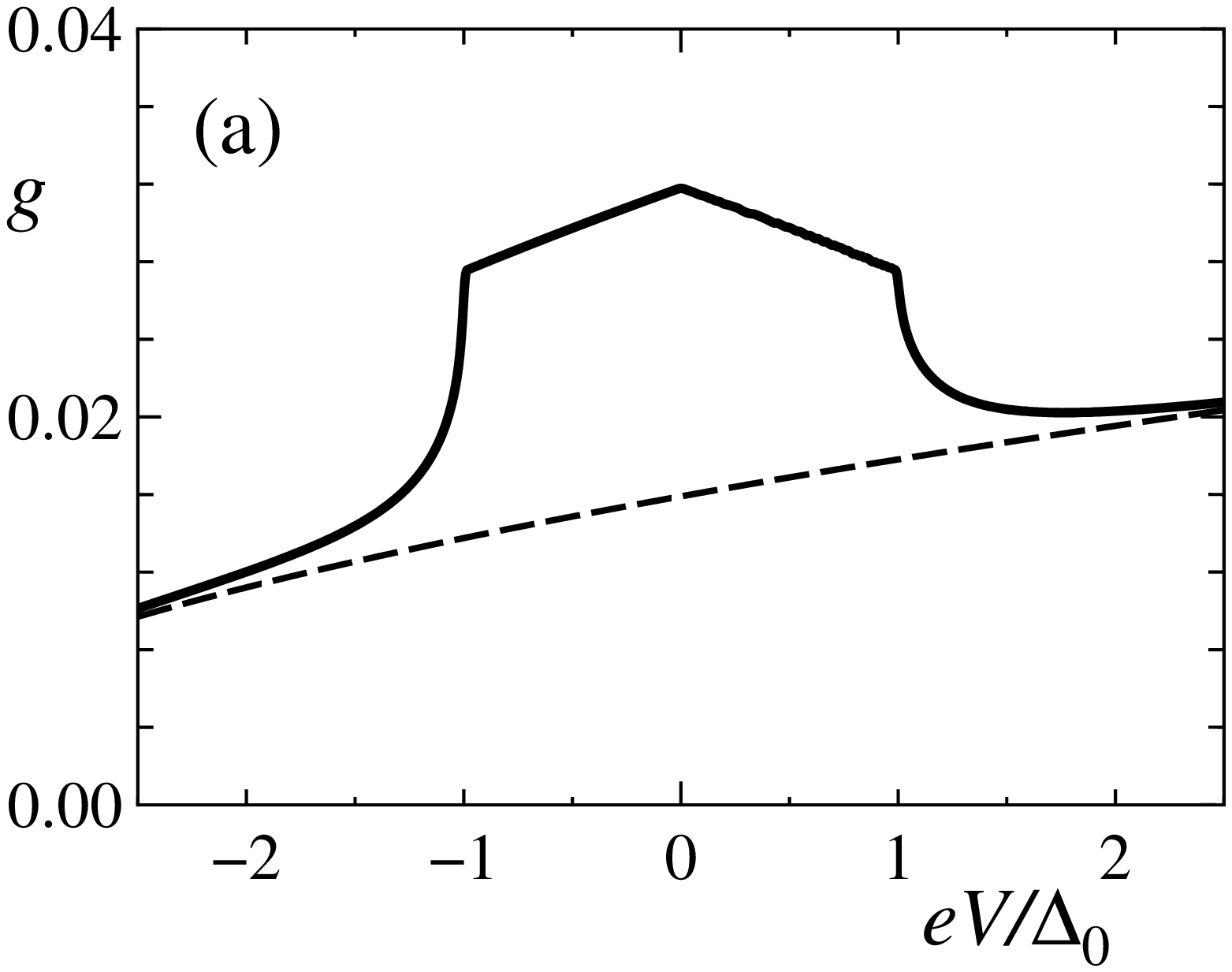}
\includegraphics[height=4.0cm]{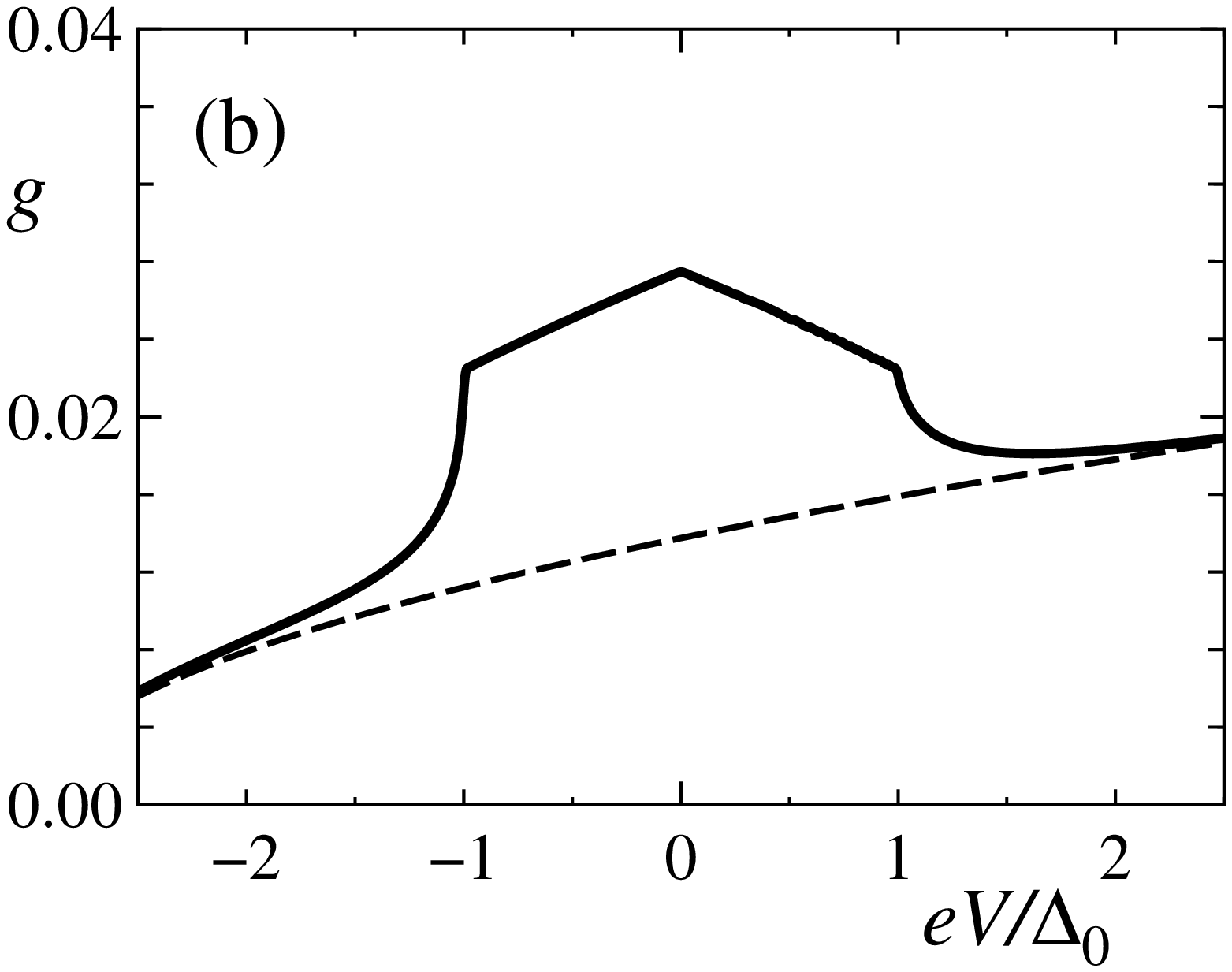}
\includegraphics[height=4.0cm]{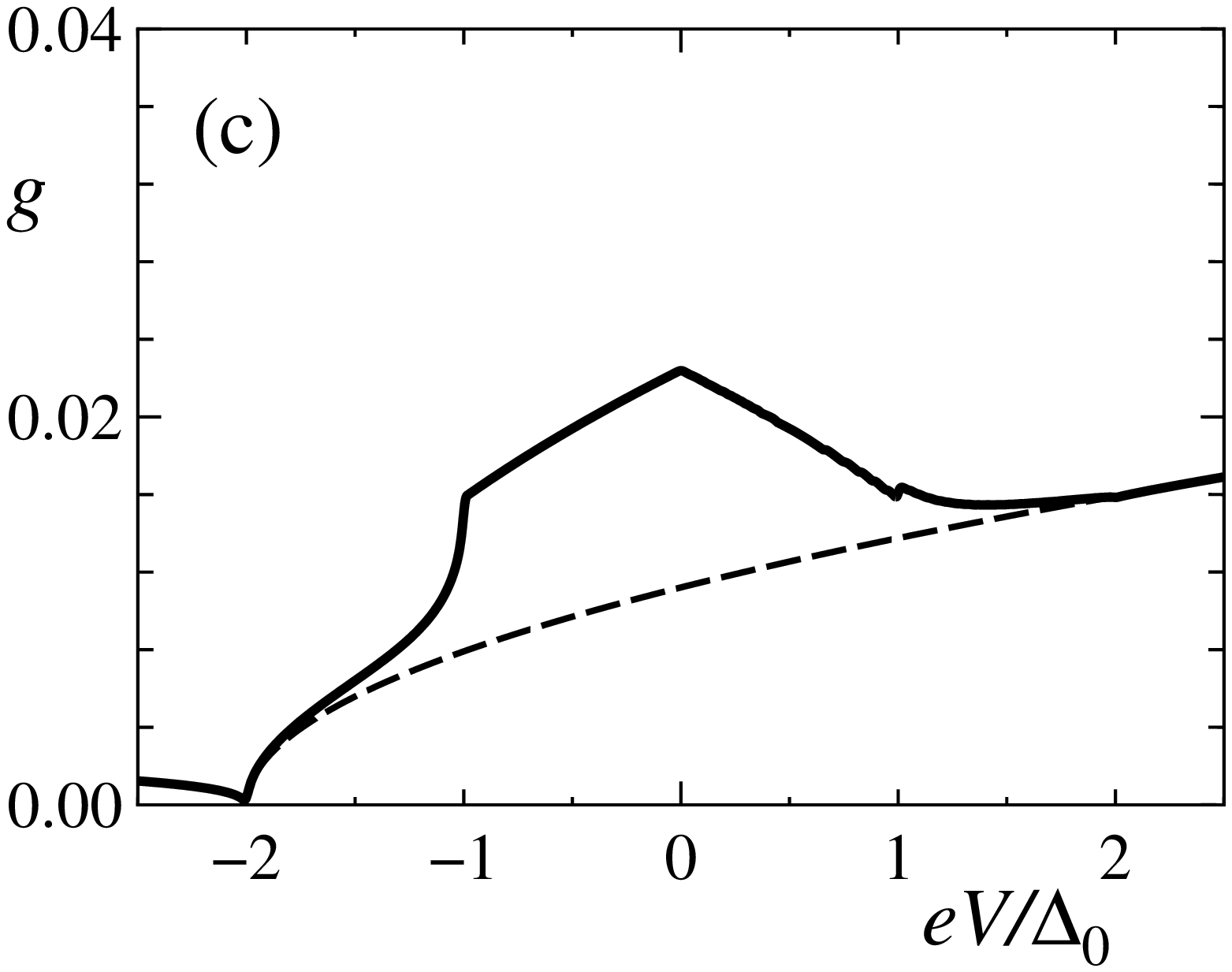}
\end{center}
\caption{
$eV$ dependence of $g_{\rm NS}(E_{\rm F}+eV)$ (solid lines)
for (a) $E_{\rm F}/\Delta_{0} = 4.0$, (b) $3.0$, and (c) $2.0$.
Dashed lines represent $g_{\rm N}(E_{\rm F}+eV)$
in the normal state (i.e., $\Delta_{0} = 0$).
}
\end{figure}

\begin{figure}[tbp]
\begin{center}
\includegraphics[height=4.0cm]{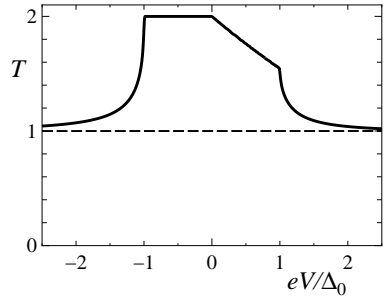}
\end{center}
\caption{
$eV$ dependence of $T_{\rm NS}(E_{\rm F}+eV)$ (solid line)
for $E_{\rm F}/\Delta_{0} = 4.0$.
The dashed line represents $T_{\rm N}(E_{\rm F}+eV)$
in the normal state (i.e., $\Delta_{0} = 0$).
}
\end{figure}

In this section, we present the numerical results of $g_{\rm NS}$,
as well as those of $g_{\rm N}$, $T_{\rm NS}$, and $T_{\rm N}$ for reference,
as a function of $eV$ for several values of $E_{\rm F}$
in the low-energy regime of $|E_{\rm F}|/\Delta_{0} \le 5$.
Our model for a bilayer graphene junction is characterized by the following parameters: $\gamma_{0}$, $\gamma_{1}$, $\Delta_{0}$, $U_{0}$, and $\lambda$.
The transfer integrals are given by
$\gamma_{0} = 3.16$ eV and $\gamma_{1} = 0.39$ eV.
Assuming that $\rm NbSe_{2}$ is deposited on a bilayer graphene sheet
as a bulk superconductor,~\cite{efetov1} we set $\Delta_{0}= 1.2$ meV.
The work functions of $\rm NbSe_{2}$ and bilayer graphene have been reported
to be $\Phi_{\rm NbSe_2} = 5.9$ eV~\cite{shimada}
and $\Phi_{\rm BG} = 4.4$ eV.~\cite{hibino}
Roughly speaking, their difference is on the order of $1$ eV;
thus, we set $U_{0} = 1.0$ eV.
A plausible value of $\lambda$ is not known, although it is believed to be
on the order of $100$ nm or much longer.
We consider the case of $\lambda = 100$ nm as the main case
as well as the supplementary
cases of $\lambda = 0.0$, $2.46$, $4.92$, $40$, and $400$ nm.
Here, $\lambda = 2.46$ ($4.92$) nm corresponds to $\lambda/a = 10$ ($20$).

\begin{figure}[tbp]
\begin{center}
\includegraphics[height=4.0cm]{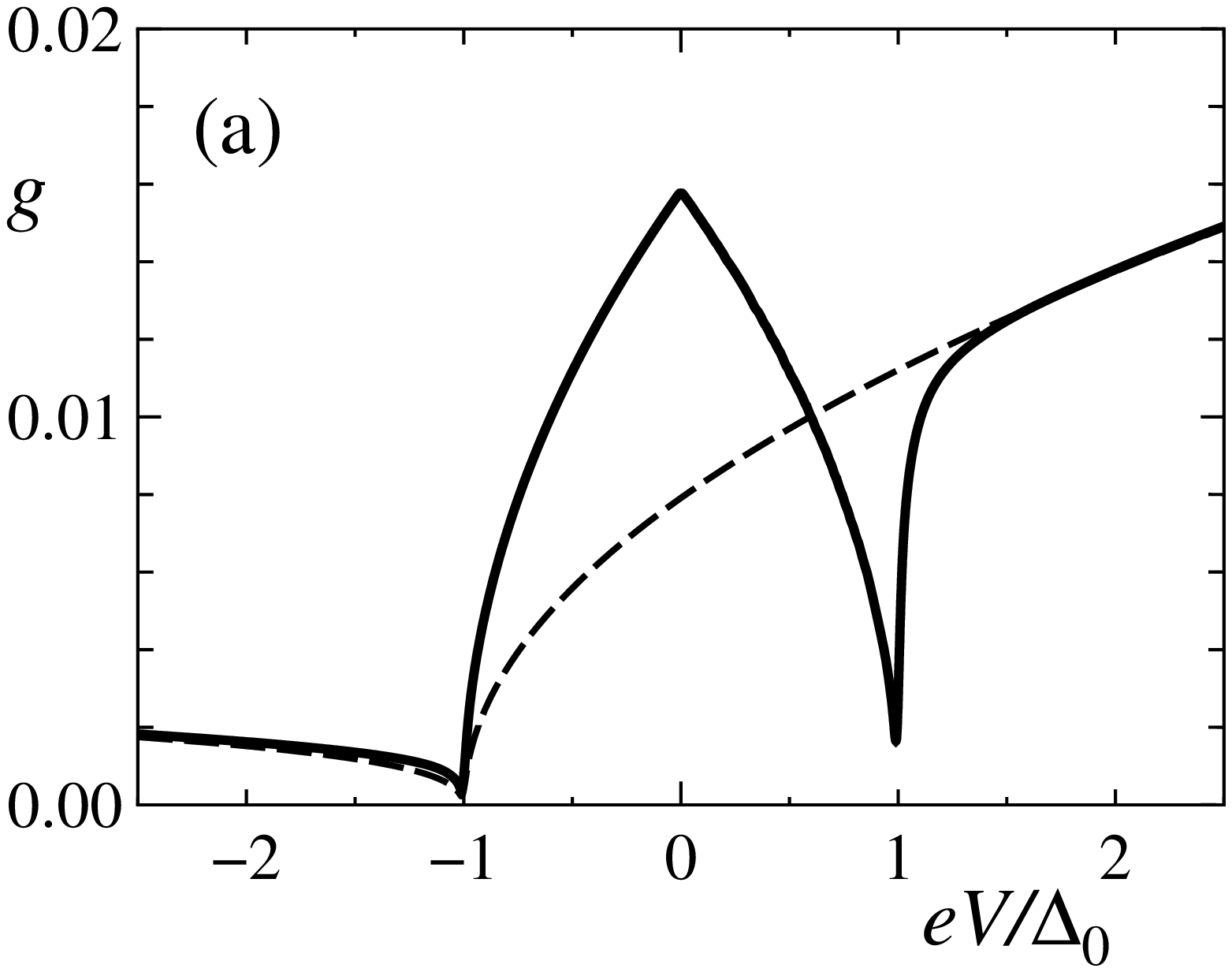}
\includegraphics[height=4.0cm]{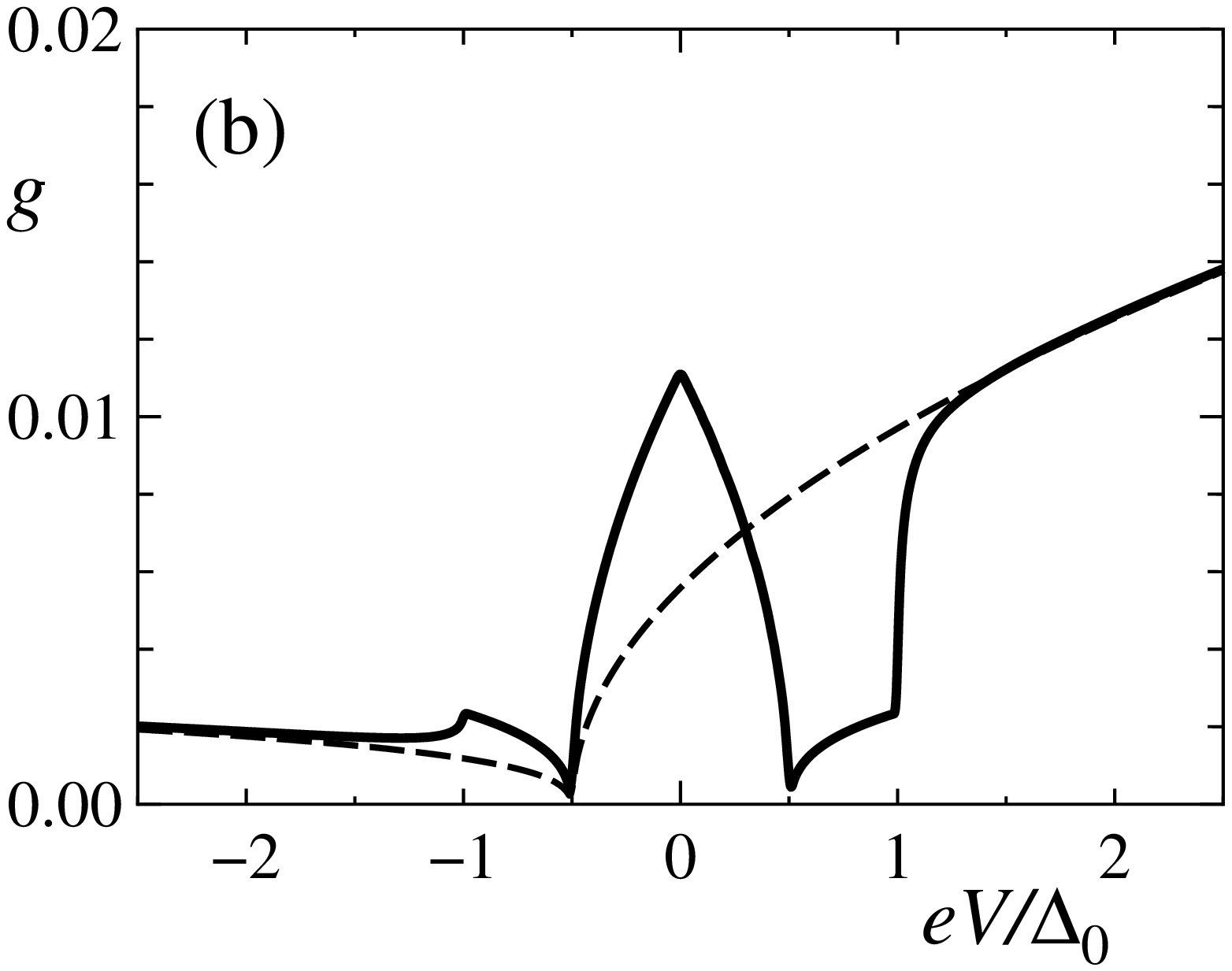}
\includegraphics[height=4.0cm]{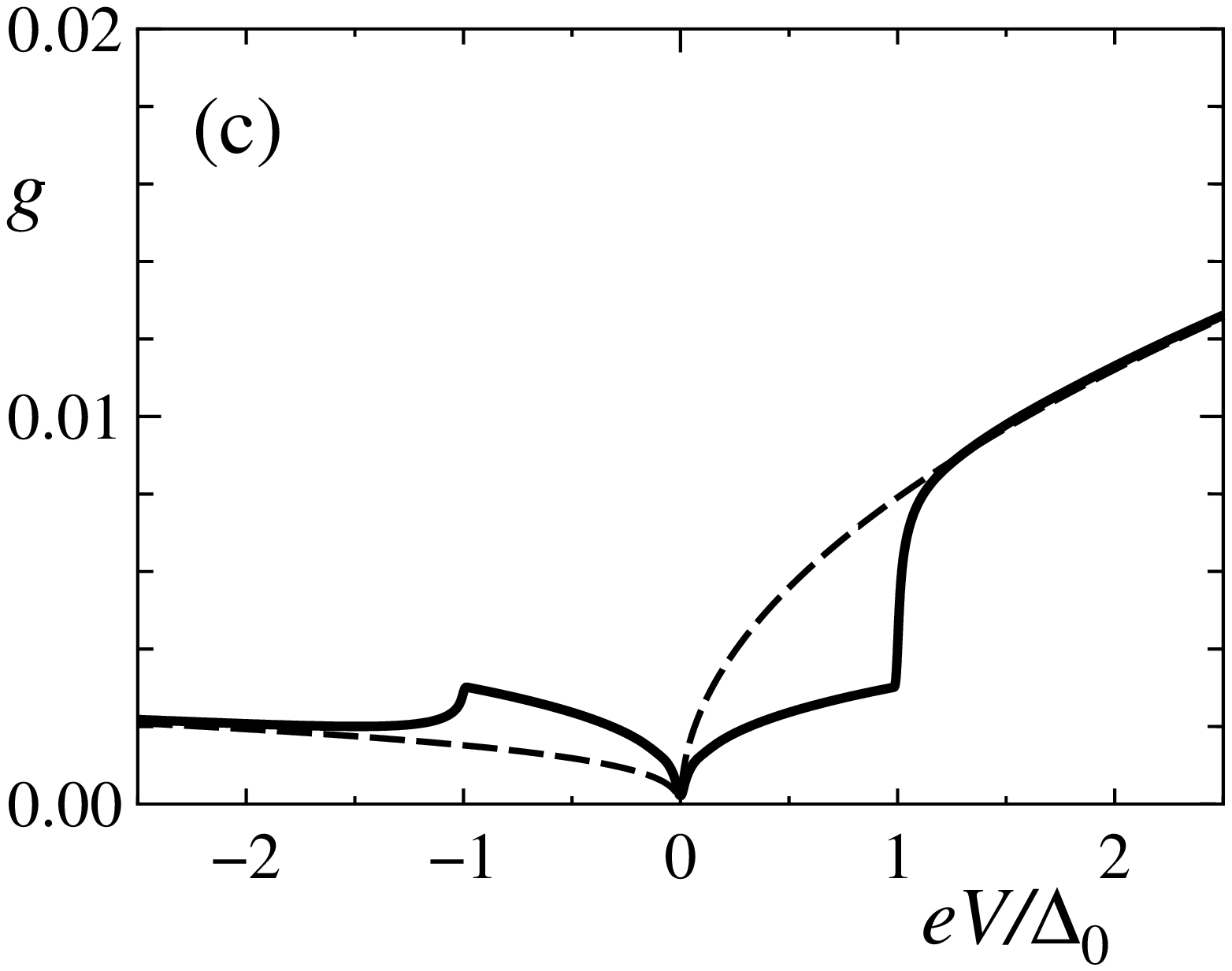}
\end{center}
\caption{
$eV$ dependence of $g_{\rm NS}(E_{\rm F}+eV)$
for (a) $E_{\rm F}/\Delta_{0} = 1.0$, (b) $0.5$, (c) $0.0$.
Dashed lines represent $g_{\rm N}(E_{\rm F}+eV)$
in the normal state (i.e., $\Delta_{0} = 0$).
}
\end{figure}

We start with the case of $\lambda = 100$ nm.
Figure~2 shows $g_{\rm NS}(E_{\rm F}+eV)$ as a function of $eV$
for $E_{\rm F}/\Delta_{0} = 4.0$, $3.0$, and $2.0$.
The excess conductance due to the AR is clearly seen in the subgap region of
$\Delta_{0} > eV > -\Delta_{0}$.
The behavior of $g_{\rm NS}$ is similar to that observed in an ordinary
NS junction composed of a normal metal and a superconductor.
A slightly cusped peak at $eV/\Delta_{0} = 0$ is accounted for
if one recognizes that $k_{\rm F}(E_{\rm F}+eV)$ with $eV > 0$ is
greater than that with $eV < 0$.
Let us consider the AR process in which an electron with energy
$E_{\rm F} + eV$ is reflected back as a hole with energy $E_{\rm F} - eV$.
Note that the transverse wave number $q$ is common to both.
However, the allowed values of $q$  are different between
electron and hole states: $|q| < k_{\rm F}(E_{\rm F}+eV)$ for an electron
while $|q| < k_{\rm F}(E_{\rm F}-eV)$ for a hole.
As the AR is a scattering process between an electron and a hole,
it can be allowed only for $q$ satisfying $|q| < q_{\rm c}$ with $q_{\rm c}
\equiv {\rm min}\{k_{\rm F}(E_{\rm F}+eV), k_{\rm F}(E_{\rm F}-eV)\}$.
This restriction results in a cusp at $eV/\Delta_{0} = 0$.
To see the AR probability in the subgap region, we show
$T_{\rm NS}(E_{\rm F}+eV)$ for $E_{\rm F}/\Delta_{0} = 4.0$ in Fig.~3,
which clearly indicates that
the AR probability is $1$ in the region of $0 > eV > -\Delta_{0}$.
In the region of $\Delta_{0} > eV > 0$,
we see that $T_{\rm NS}(E_{\rm F}+eV)$ decreases with increasing $eV$.
This reflects the fact that an electron with $q$ satisfying
$k_{\rm F}(E_{\rm F}+eV) > |q| > k_{\rm F}(E_{\rm F}-eV)$
cannot be reflected as a hole owing to the restriction for the hole state
[i.e., $|q| < k_{\rm F}(E_{\rm F}-eV)$].

\begin{figure}[btp]
\begin{center}
\includegraphics[height=5.0cm]{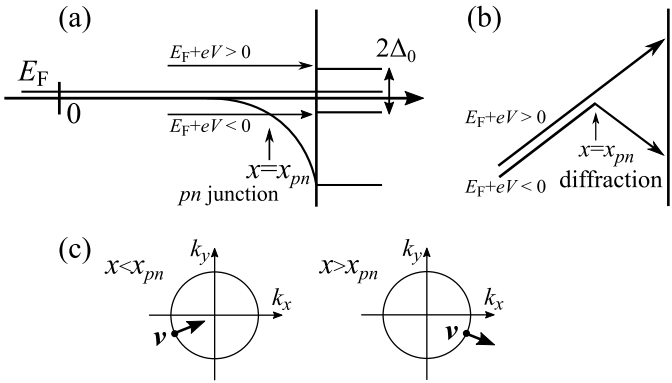}
\end{center}
\caption{
(a) An electron with energy $E_{\rm F}+eV<0$ passes through
a naturally formed $pn$ junction,
while such a $pn$ junction is not present for an electron
with energy $E_{\rm F}+eV>0$.
(b) An electron with energy $E_{\rm F}+eV<0$ is diffracted
at the $pn$ junction.
The curvature of trajectories, which arises from the $x$ dependence of
the longitudinal wave number $k_{x}$, is ignored here and hereafter.
(c) The diffraction of an electron with a negative energy is explained
as follows.
In the region of $x<x_{pn}$ where the electron is in the valence band,
the group velocity $\mib{v}$ of a state with $\mib{k}=(k_{x},k_{y})$
is antiparallel to $\mib{k}$,
while $\mib{v}$ is parallel to $\mib{k}$ in the region of $x>x_{pn}$
where the electron is in the conduction band.
This combined with the conservation of $k_{y}$
results in the diffraction of an electron trajectory as shown in (b).
}
\end{figure}

\begin{figure}[btp]
\begin{center}
\includegraphics[height=4.0cm]{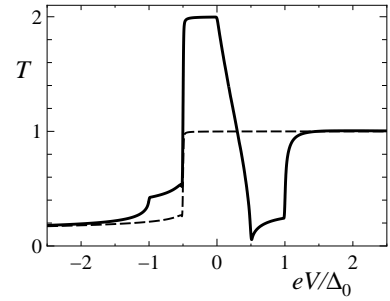}
\end{center}
\caption{
$eV$ dependence of $T_{\rm NS}(E_{\rm F}+eV)$ (solid line)
and $T_{\rm N}(E_{\rm F}+eV)$ (dashed line)
for $E_{\rm F}/\Delta_{0} = 0.5$.
}
\end{figure}

\begin{figure}[btp]
\begin{center}
\includegraphics[height=2.5cm]{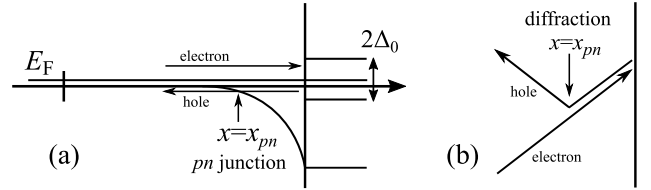}
\end{center}
\caption{
(a) When $\Delta_{0} > E_{\rm F} + eV > 0$ and
$0 > E_{\rm F} - eV > -\Delta_{0}$, only a reflected hole
passes through a $pn$ junction.
(b) A reflected hole is diffracted at the $pn$ junction,
resulting in the specular nature of the AR.
}
\end{figure}

\begin{figure}[tbp]
\begin{center}
\includegraphics[height=4.0cm]{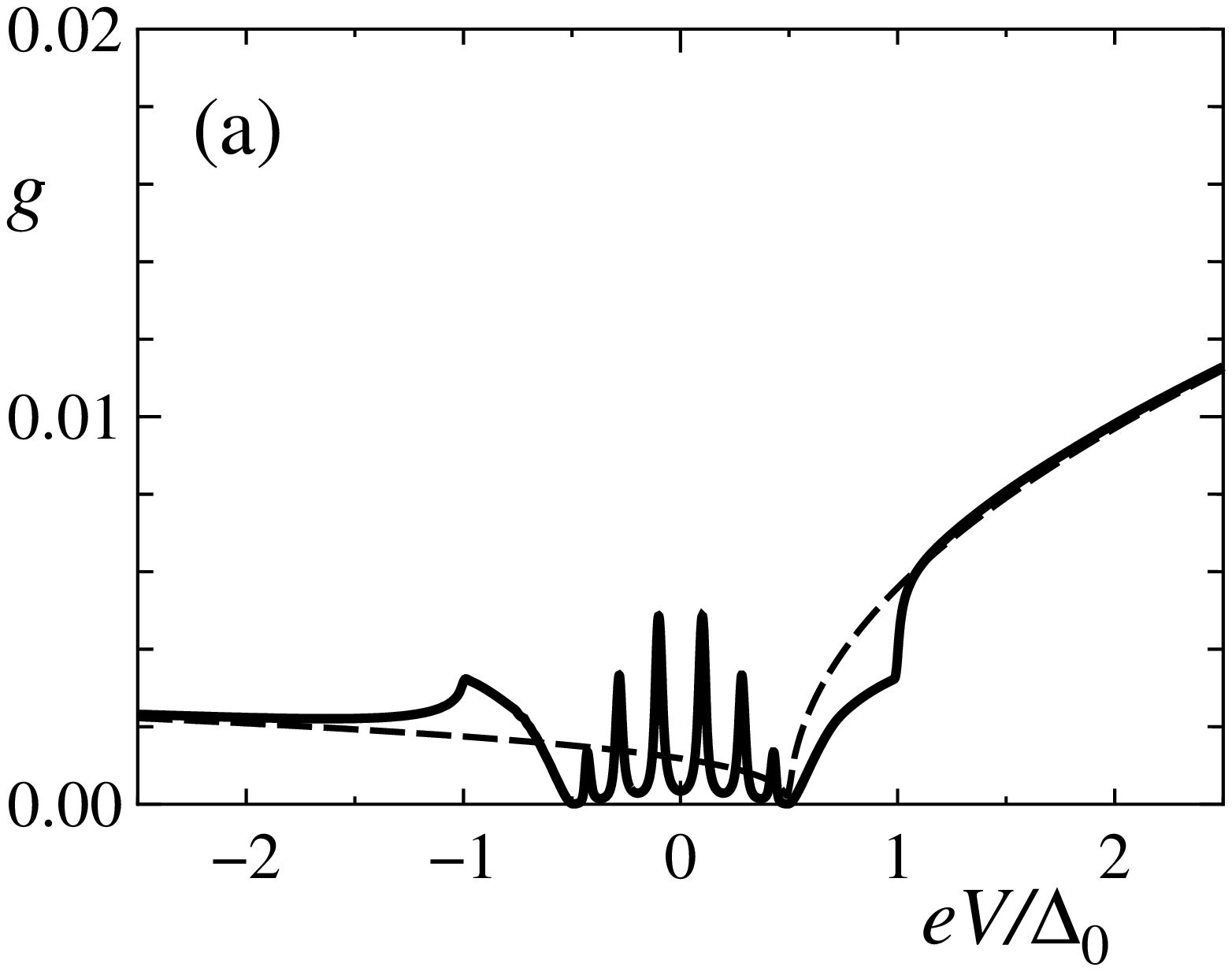}
\includegraphics[height=4.0cm]{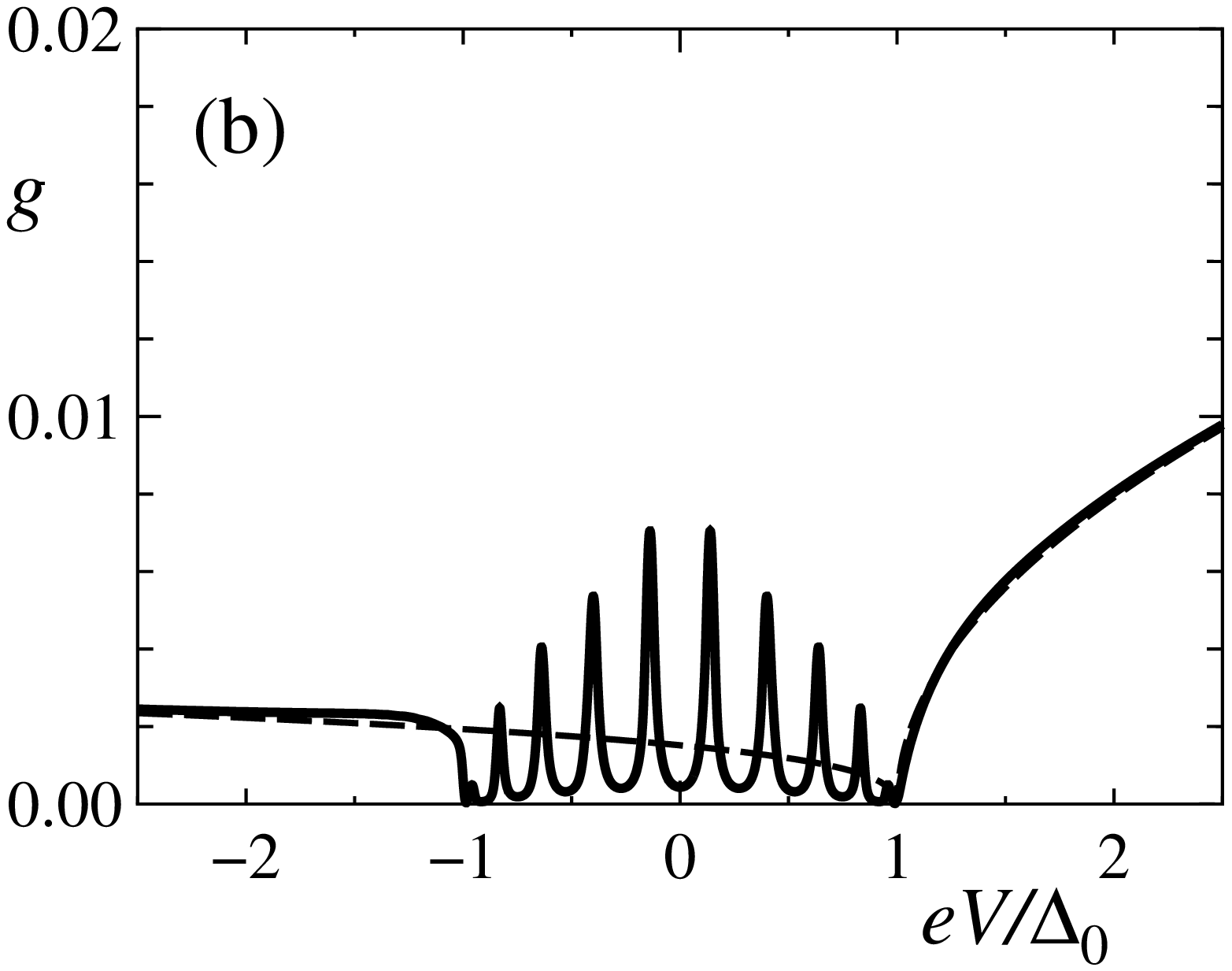}
\includegraphics[height=4.0cm]{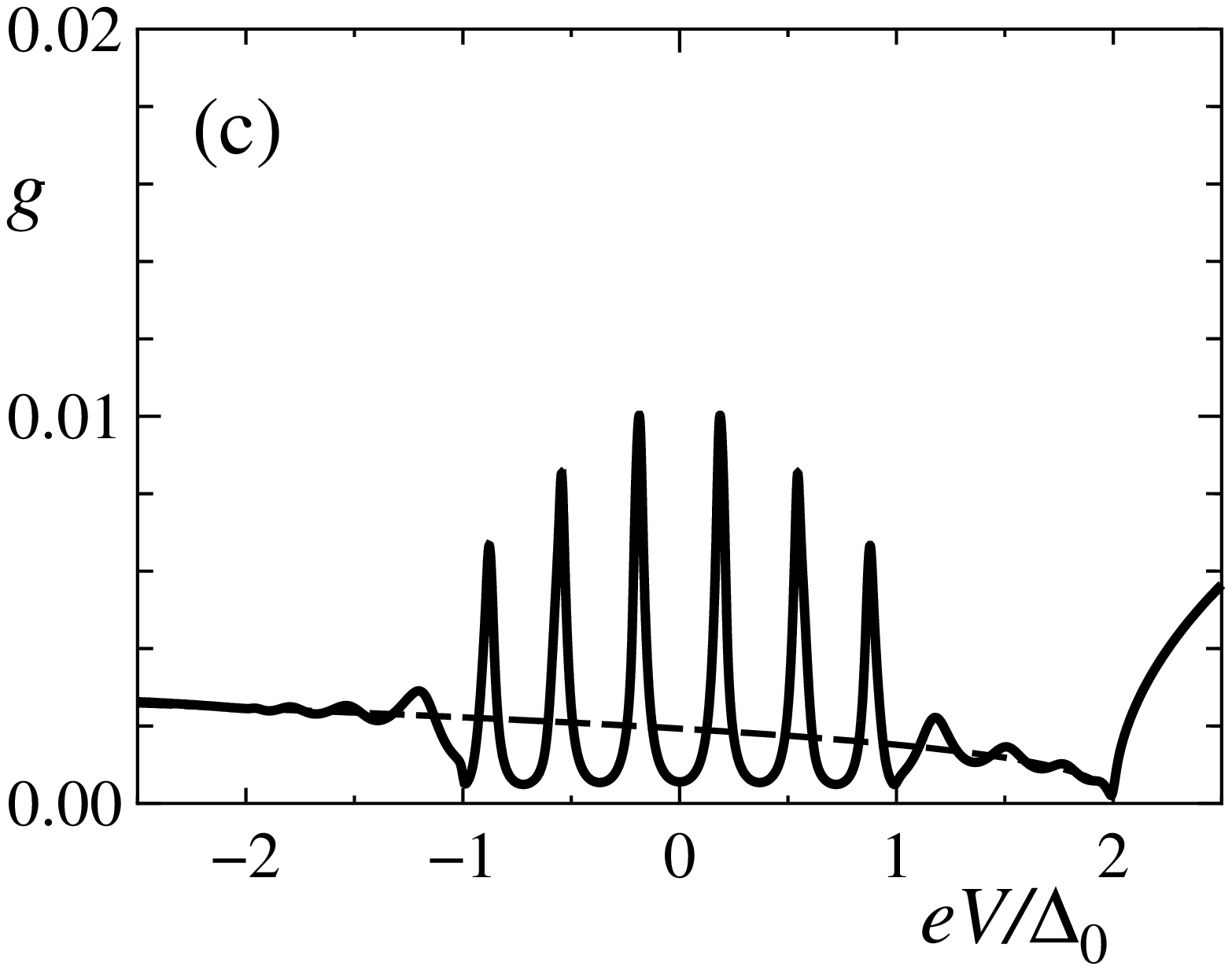}
\includegraphics[height=4.0cm]{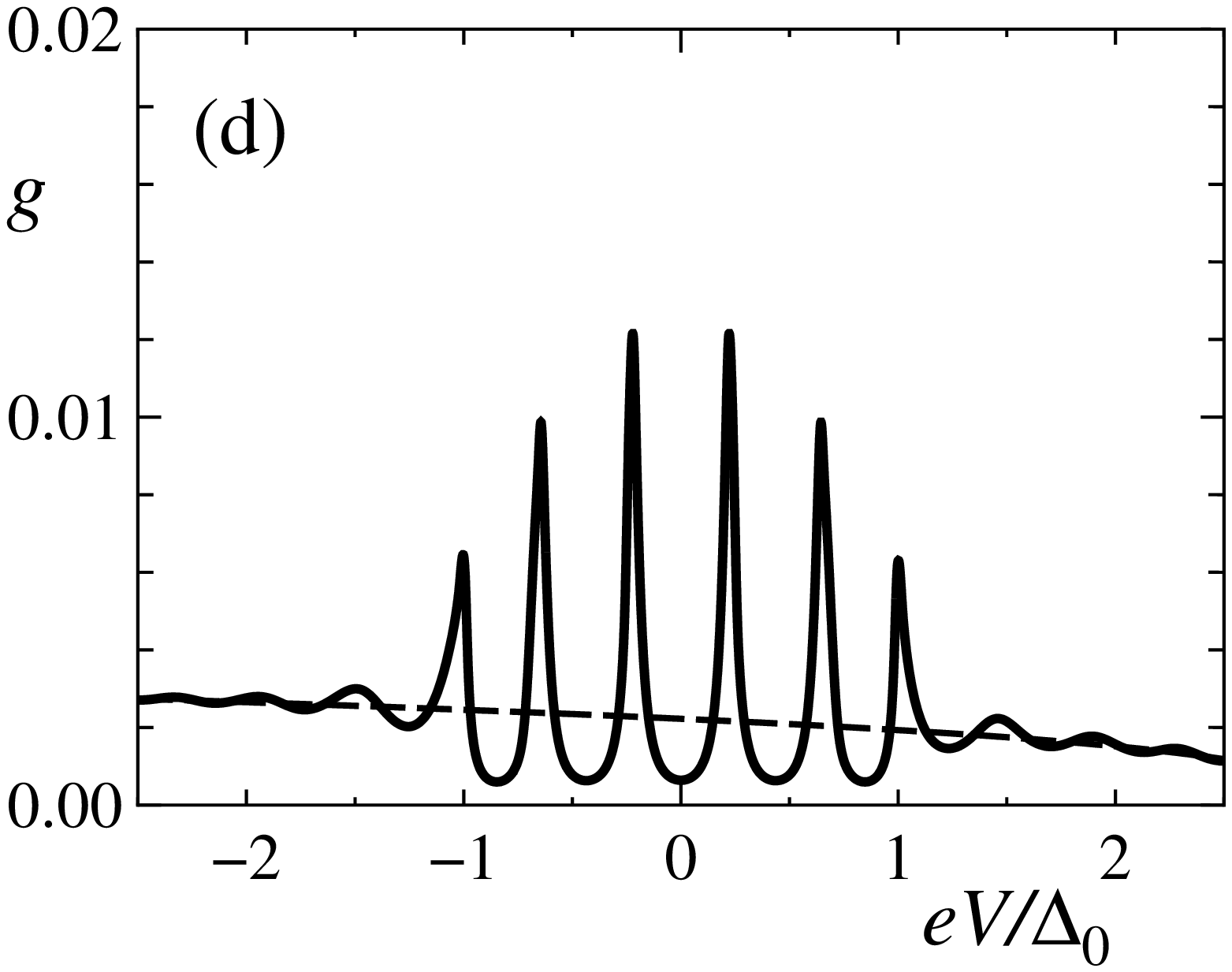}
\end{center}
\caption{
$eV$ dependence of $g_{\rm NS}(E_{\rm F}+eV)$ (solid lines)
for (a) $E_{\rm F}/\Delta_{0} = -0.5$,
(b) $-1.0$, (c) $-2.0$, and (d) $-3.0$.
Dashed lines represent $g_{\rm N}(E_{\rm F}+eV)$
in the normal state (i.e., $\Delta_{0} = 0$).
}
\end{figure}

Figure~4 shows $g_{\rm NS}(E_{\rm F}+eV)$ for
$E_{\rm F}/\Delta_{0} = 1.0$, $0.5$, and $0.0$.
The $V$ dependence in the subgap region is slightly complicated and depends
on $E_{\rm F}/\Delta_{0}$ but has a common feature that
$g_{\rm NS}(E_{\rm F}+eV)$ vanishes at $eV = - E_{\rm F}$
as $k_{\rm F}(E_{\rm F}+eV)$ becomes zero.
It should be noted here that the AR takes place in a specular manner
in the region of $\Delta_{0} > e|V| > E_{\rm F}$.
Before considering the origin of specular AR,
we pay attention to an interesting feature of $g_{\rm NS}$ revealed
in the above gap region of $e|V| > \Delta_{0}$,
where $g_{\rm NS}(E_{\rm F}+eV)$ with $eV > \Delta_{0}$
is much greater than that with $eV < -\Delta_{0}$.
As $g_{\rm NS}$ is nearly equal to $g_{\rm N}$,
this asymmetry is not related to the AR.
Indeed, it is caused by a $pn$ junction naturally formed
in the uncovered region [see Fig.~5(a)].
Let us consider an electron injected in the right direction
from the asymptotic region of $x \le 0$.
If its energy $E$ is negative, the electron passes through the crossing point
($x = x_{pn}$) at which $E$ coincides with the CNP.
This electron is in the valence band on the left-hand side of
the crossing point (i.e., $x < x_{pn}$) while it moves to the conduction band
on the right-hand side (i.e., $x > x_{pn}$); thus, its behavior near
the crossing point is equivalent to
that in a smooth $pn$ junction in graphene.~\cite{cheianov1,cheianov2}
This $pn$ junction reduces the transmission probability of an electron.
As a result, the transmission probability of an incident electron
is reduced only when $E_{\rm F}+eV < 0$.
This accounts for the asymmetry observed above.
Note also that a $pn$ junction causes
the diffraction of an electron [see Fig.~5(b)],
which is the origin of specular AR as we discuss below.
The effect of a $pn$ junction is most clearly observed
in $T_{\rm N}(E_{\rm F}+eV)$.
Figure~6 shows $T_{\rm N}$ as well as $T_{\rm NS}$
at $E_{\rm F}/\Delta_{0} = 0.5$, which indicates that the transmission
probability in the normal state is $1$ in the region of $E_{\rm F}+eV > 0$
while it is significantly reduced in the region of $E_{\rm F}+eV < 0$.

Let us return to specular AR.
When $\Delta_{0} > |E_{\rm F}|$, the AR takes place in a specular manner
in the region of $\Delta_{0} > e|V| > |E_{\rm F}|$
despite the fact that both an incident electron and a reflected hole are
in the conduction band near the NS interface [see Fig.~7(a)].
An explanation of specular AR is schematically given in Fig.~7(b)
for the case of $\Delta_{0} > eV > |E_{\rm F}|$.
The specular nature is caused by the diffraction of a hole
due to a $pn$ junction.
Contrastingly, it is caused by the diffraction of an incident electron
in the case of $\Delta_{0} > -eV > |E_{\rm F}|$ (figure not shown).
From Fig.~4(b), we observe that the differential conductance is smaller
in the specular region of $\Delta_{0} > e|V| > E_{\rm F}$ than
in the retro region of $E_{\rm F} > e|V|$.
This is attributed to the simple fact that either an electron or a hole
involved in specular AR passes through a $pn$ junction,
which reduces the AR probability.

\begin{figure}[bp]
\begin{center}
\includegraphics[height=2.5cm]{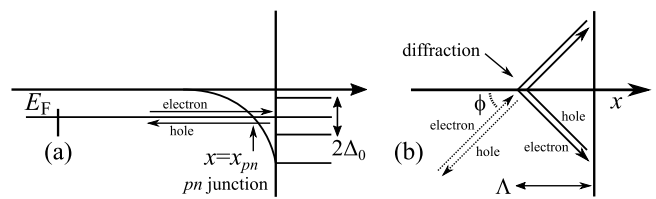}
\end{center}
\caption{(a) When $E_{\rm F} + eV < 0$ and $E_{\rm F} - eV < 0$,
both an incident electron and a reflected hole pass through a $pn$ junction.
(b) Trajectories of an electron and a hole composing quasi-bound
states (solid lines) near the NS interface.
Dotted lines represent trajectories of
an incident electron and a reflected hole.
}
\end{figure}

We turn to the case of $E_{\rm F} < 0$, in which a multiple peak structure
manifests itself for $g_{\rm NS}(E_{\rm F}+eV)$ as shown in Fig.~8
for $E_{\rm F}/\Delta_{0} = -0.5$, $-1.0$, $-2.0$, and $-3.0$.
This structure is explained as follows.
Both the energy $E_{\rm F}+eV$ of an incident electron and
the energy $E_{\rm F}-eV$ of a reflected hole are negative
in the region of $|E_{\rm F}| > e|V|$.
If $E_{\rm F}+eV < 0$ and $E_{\rm F}-eV < 0$, a $pn$ junction formed
in the uncovered region affects both of them [see Fig.~9(a)],
in contrast to the case of $\Delta_{0} > E_{\rm F} > 0$
where only one of them is affected.
The $pn$ junction plays the role of a weak barrier for an electron and a hole,
creating a quasiparticle weakly confined
between the NS interface and the $pn$ junction.
Such a quasi-bound state gives rise to
a resonant peak in the differential conductance.
In other words, the AR probability is enhanced
by the constructive interference of multiple scattering processes.
Indeed, we observe such peaks in Fig.~8 in the region of
$e|V| < {\rm min}\{\Delta_{0},|E_{\rm F}|\}$.
We expect that the spacing between two neighboring peaks is mainly
determined by the screening length $\lambda$
over which a quasiparticle is weakly confined
(see the discussion of Fig.~12 given below).

\begin{figure}[bpt]
\begin{center}
\includegraphics[height=3.3cm]{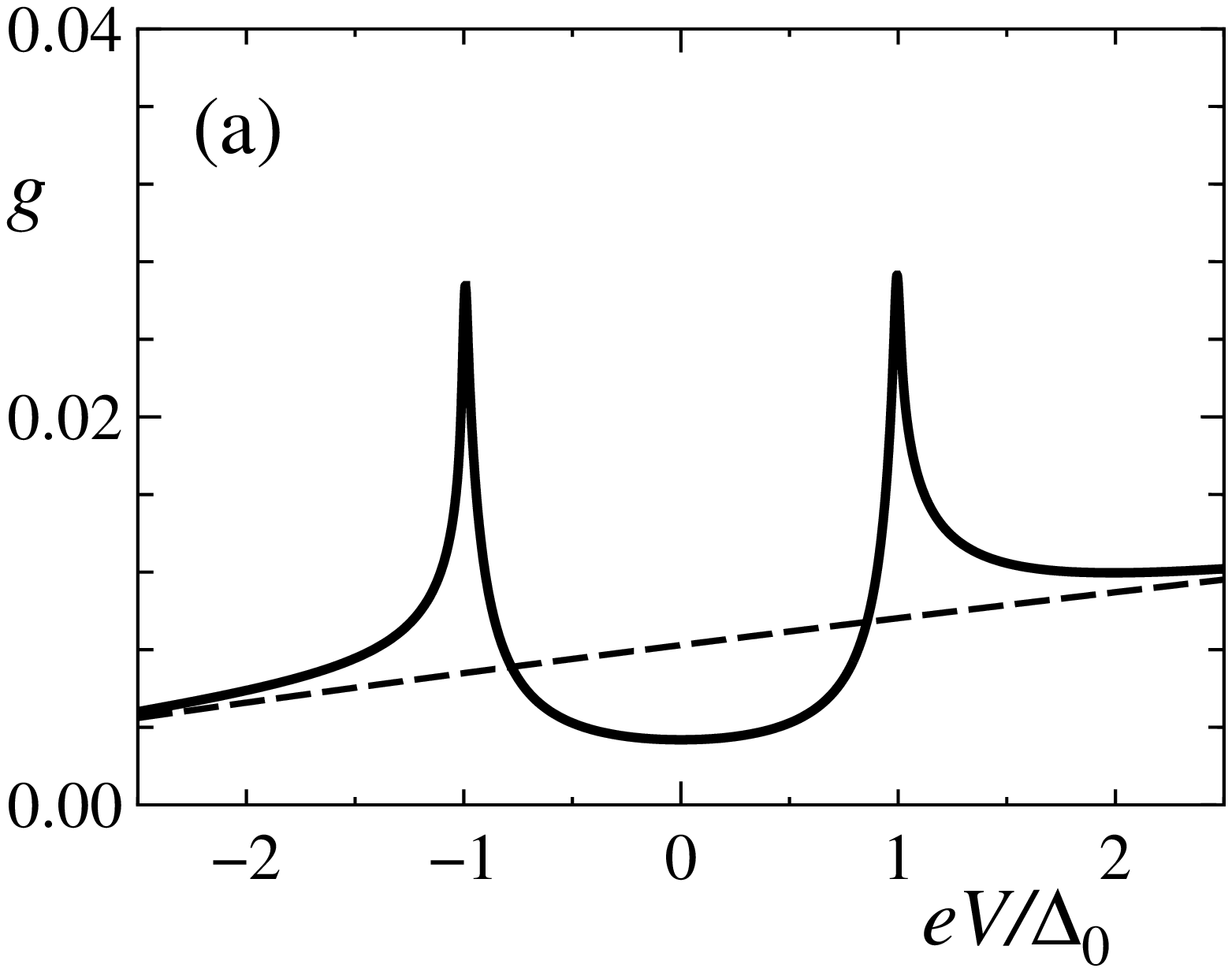}
\includegraphics[height=3.3cm]{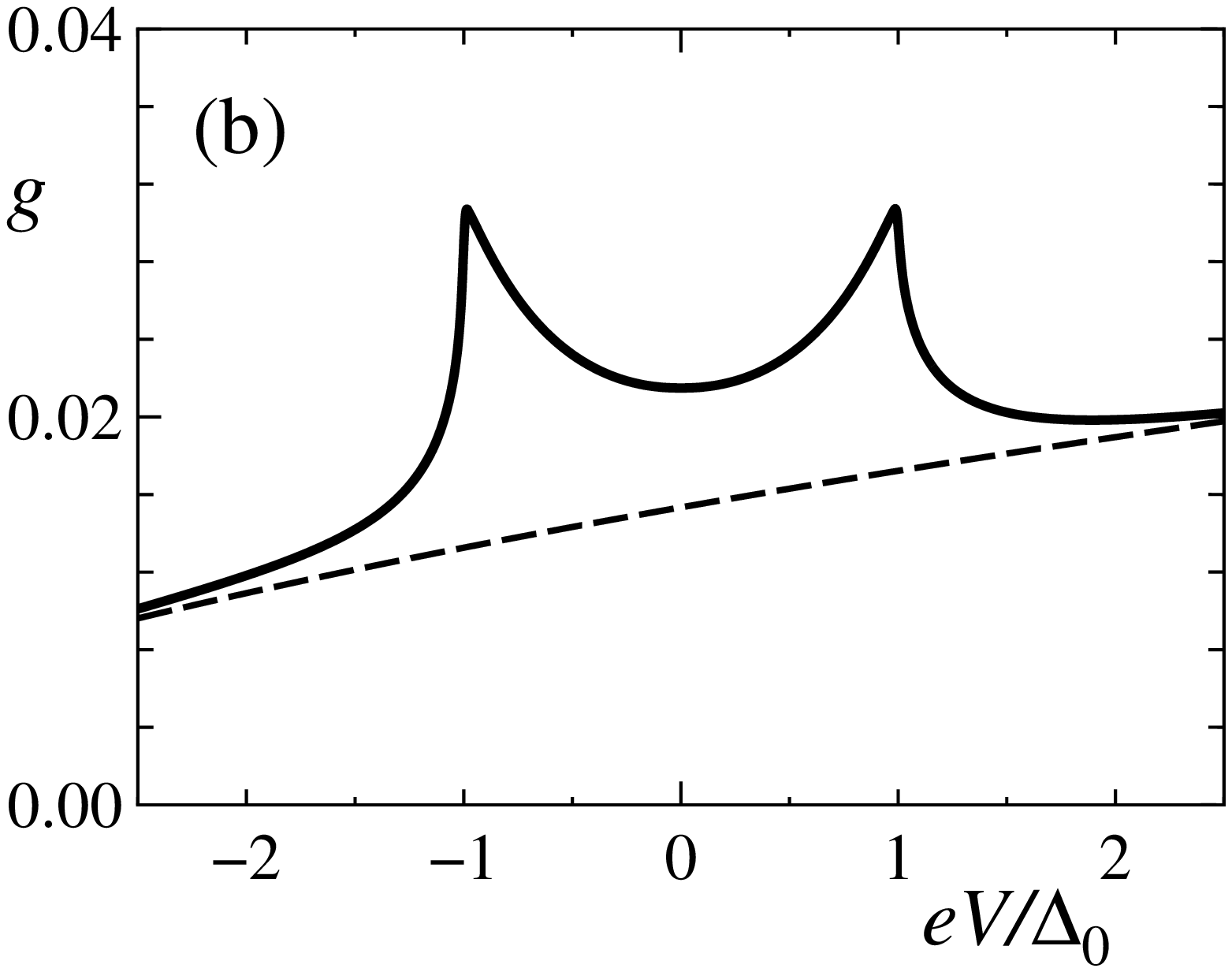}
\includegraphics[height=3.3cm]{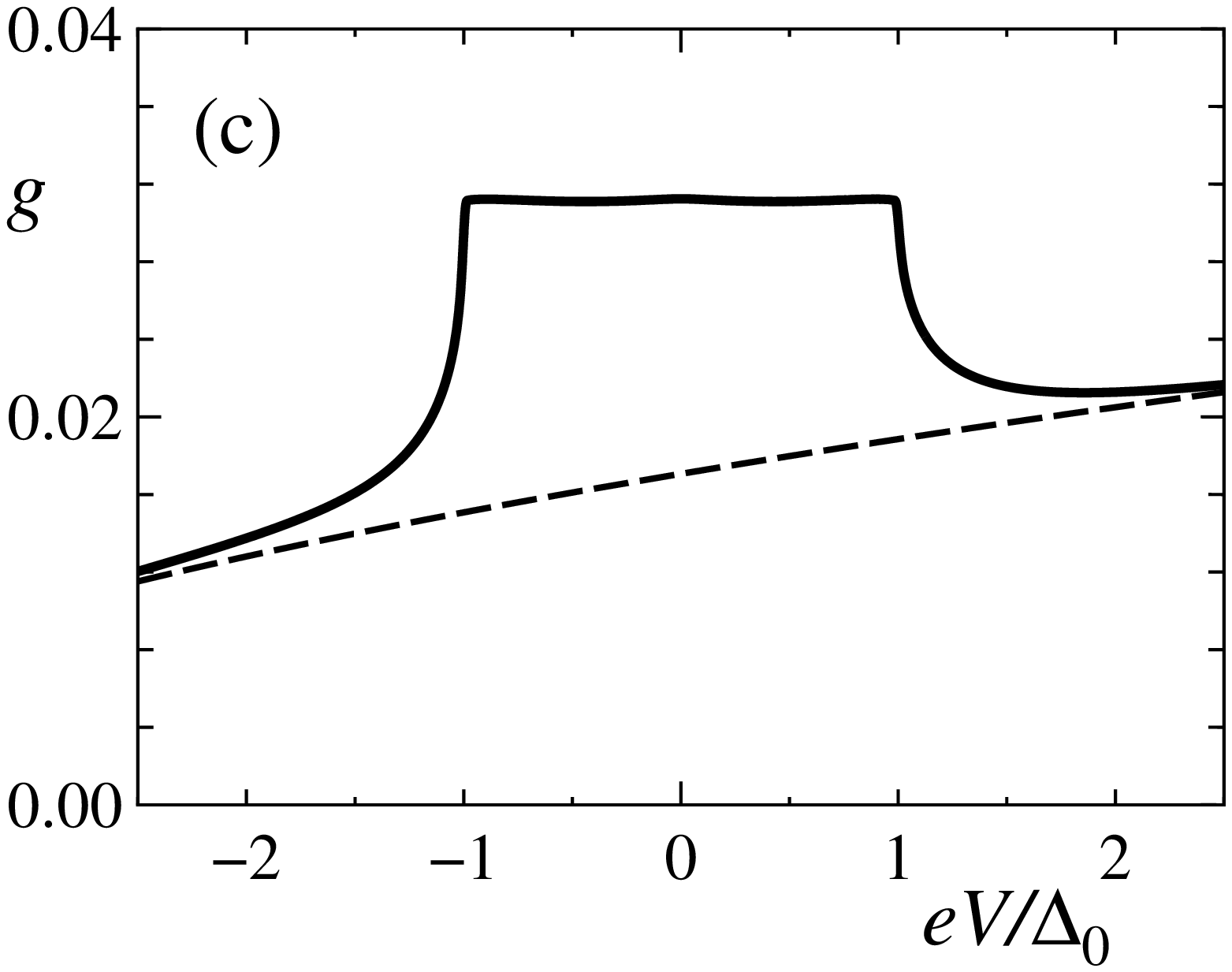}
\includegraphics[height=3.3cm]{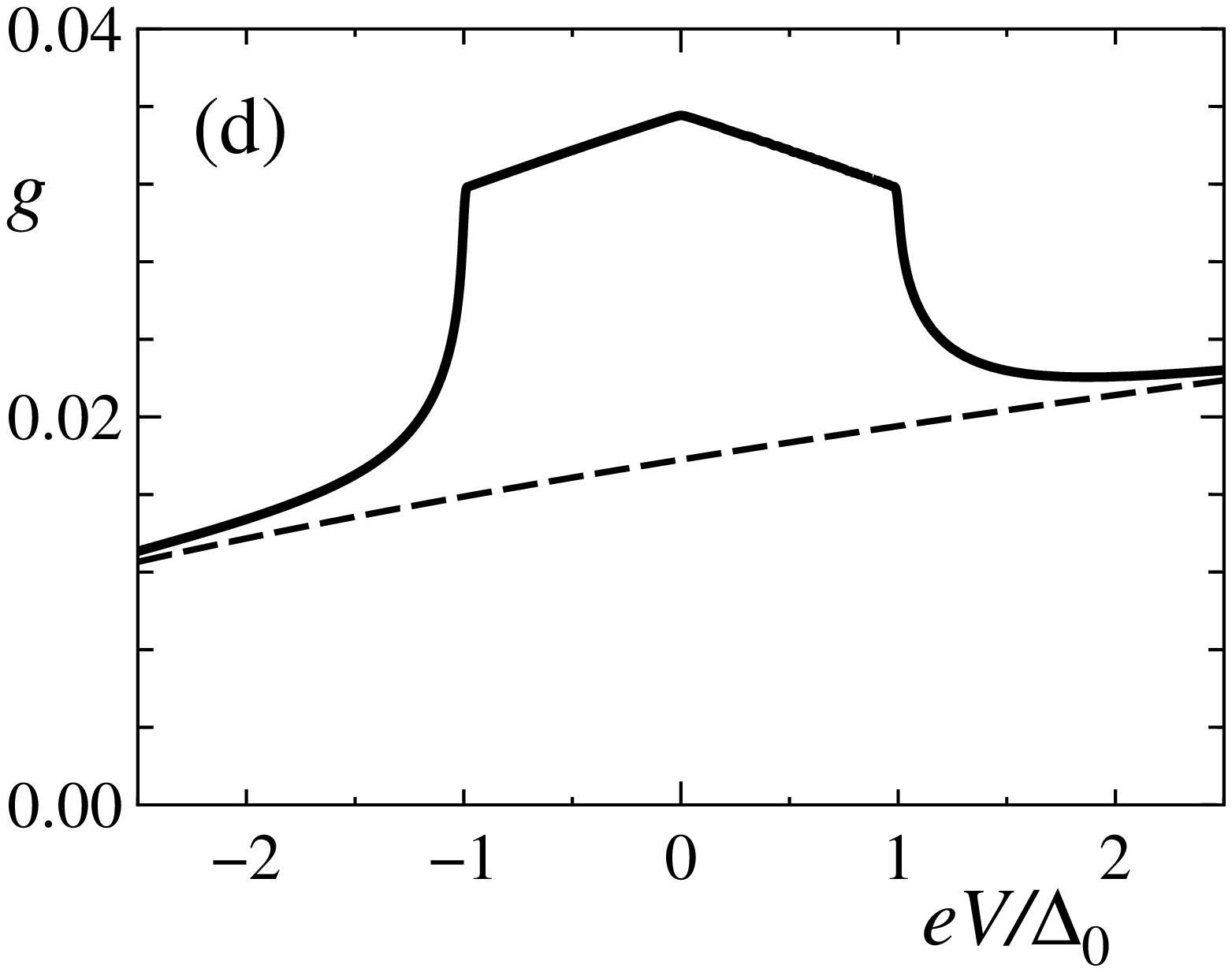}
\includegraphics[height=3.3cm]{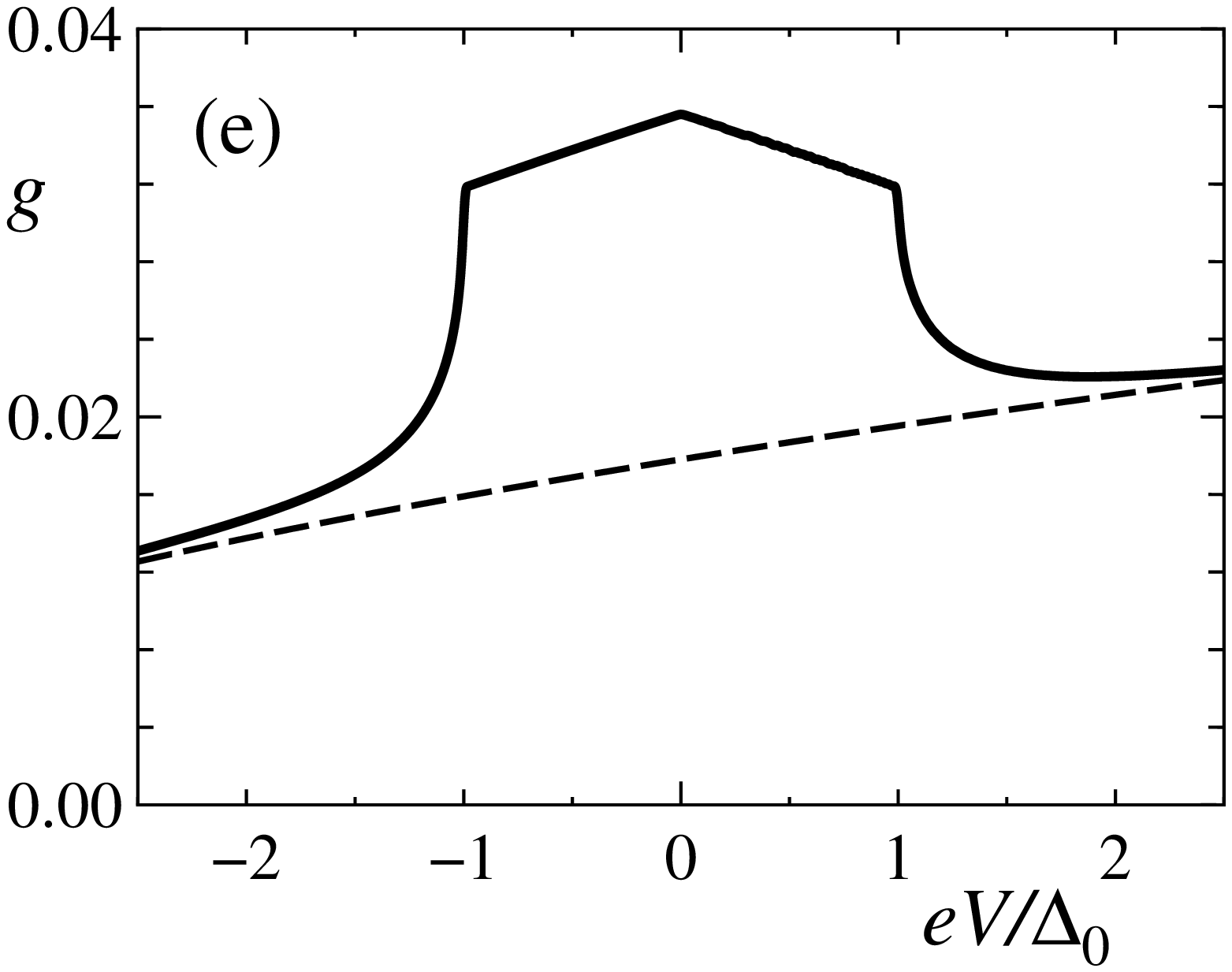}
\includegraphics[height=3.3cm]{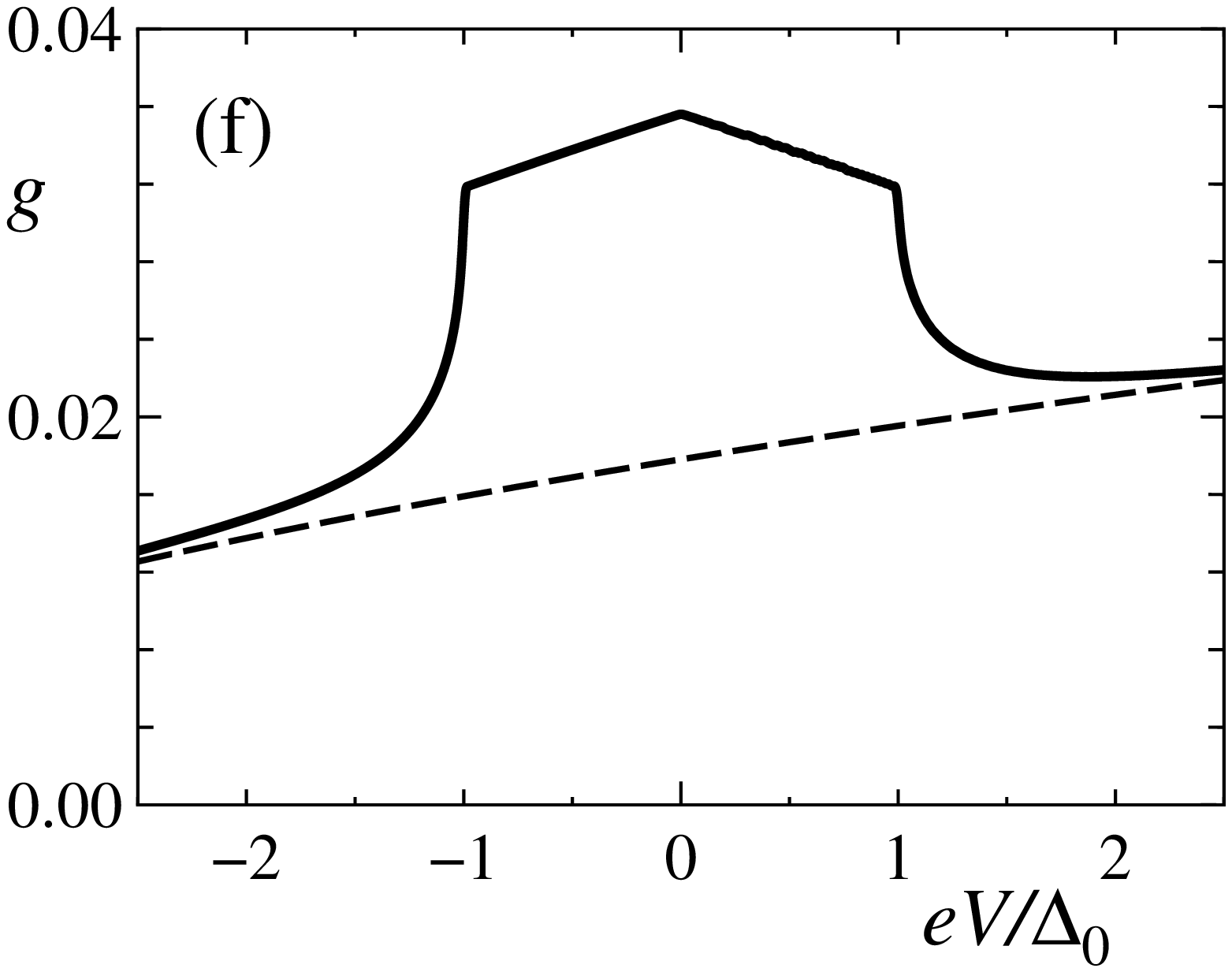}
\end{center}
\caption{
$eV$ dependence of $g_{\rm NS}(E_{\rm F}+eV)$
at $E_{\rm F}/\Delta_{0} = 5.0$ with (a) $\lambda = 0.0$, (b) $2.46$,
(c) $4.92$, (d) $40$, (e) $100$, and (f) $400$ nm.
}
\end{figure}

\begin{figure}[tbp]
\begin{center}
\includegraphics[height=3.3cm]{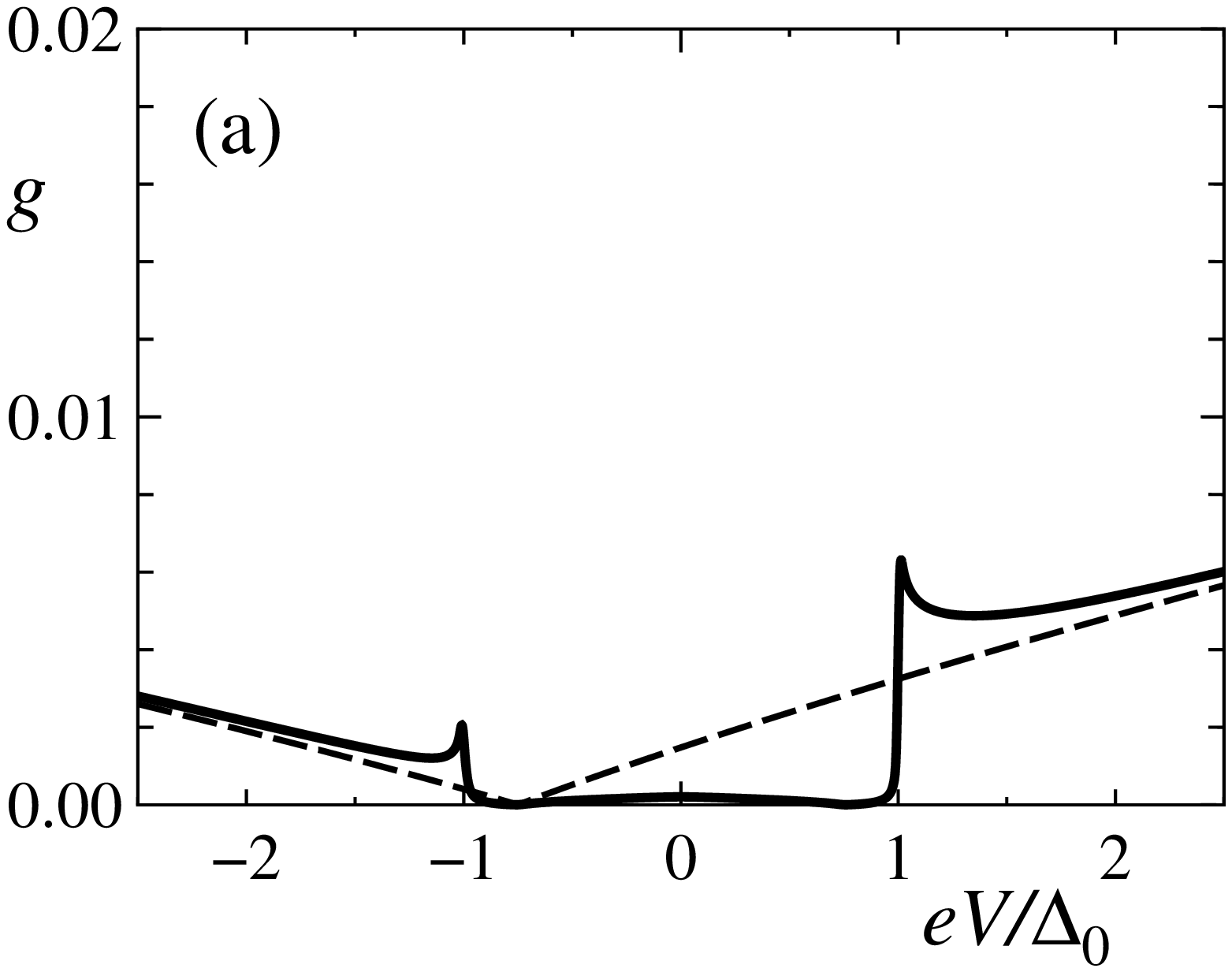}
\includegraphics[height=3.3cm]{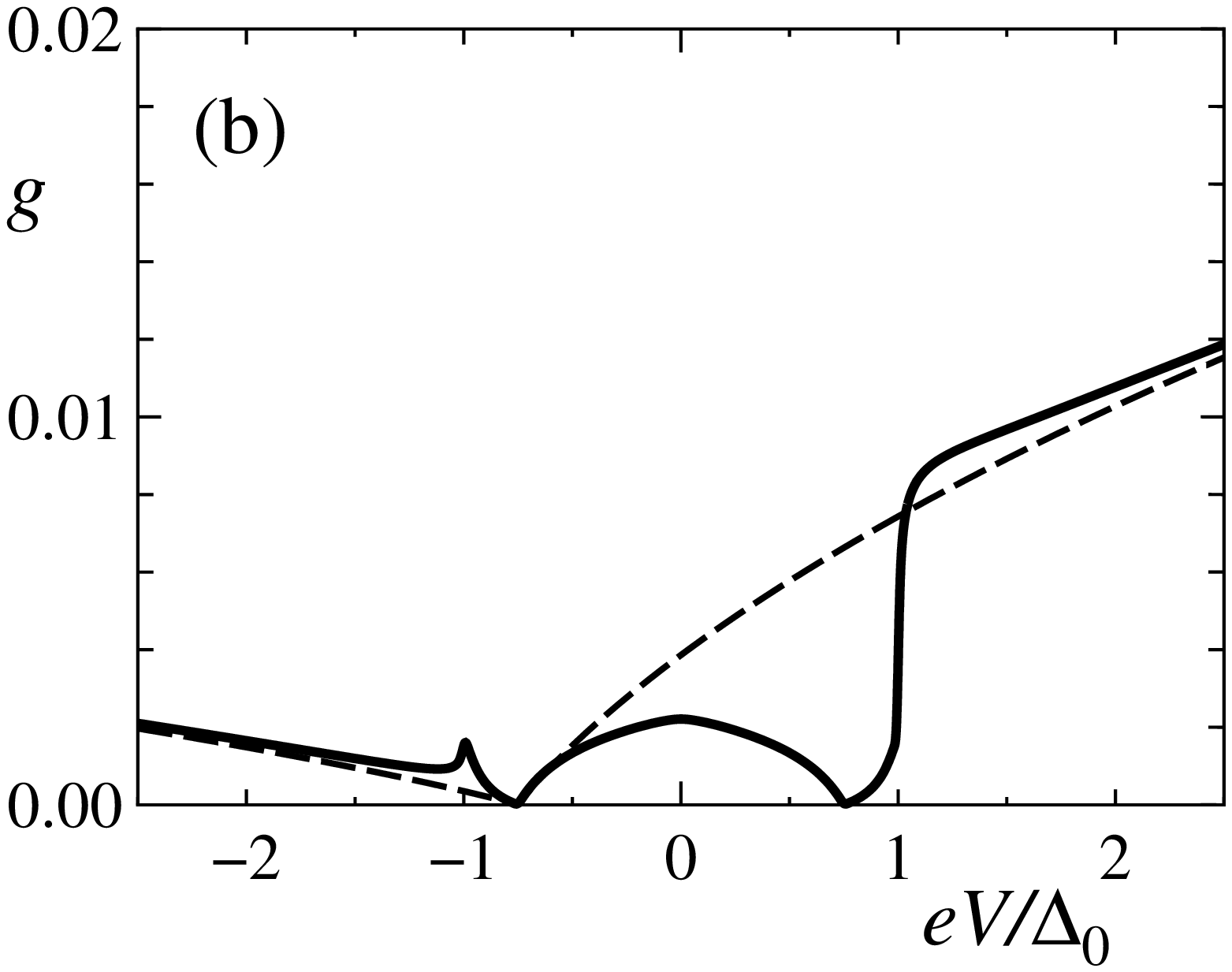}
\includegraphics[height=3.3cm]{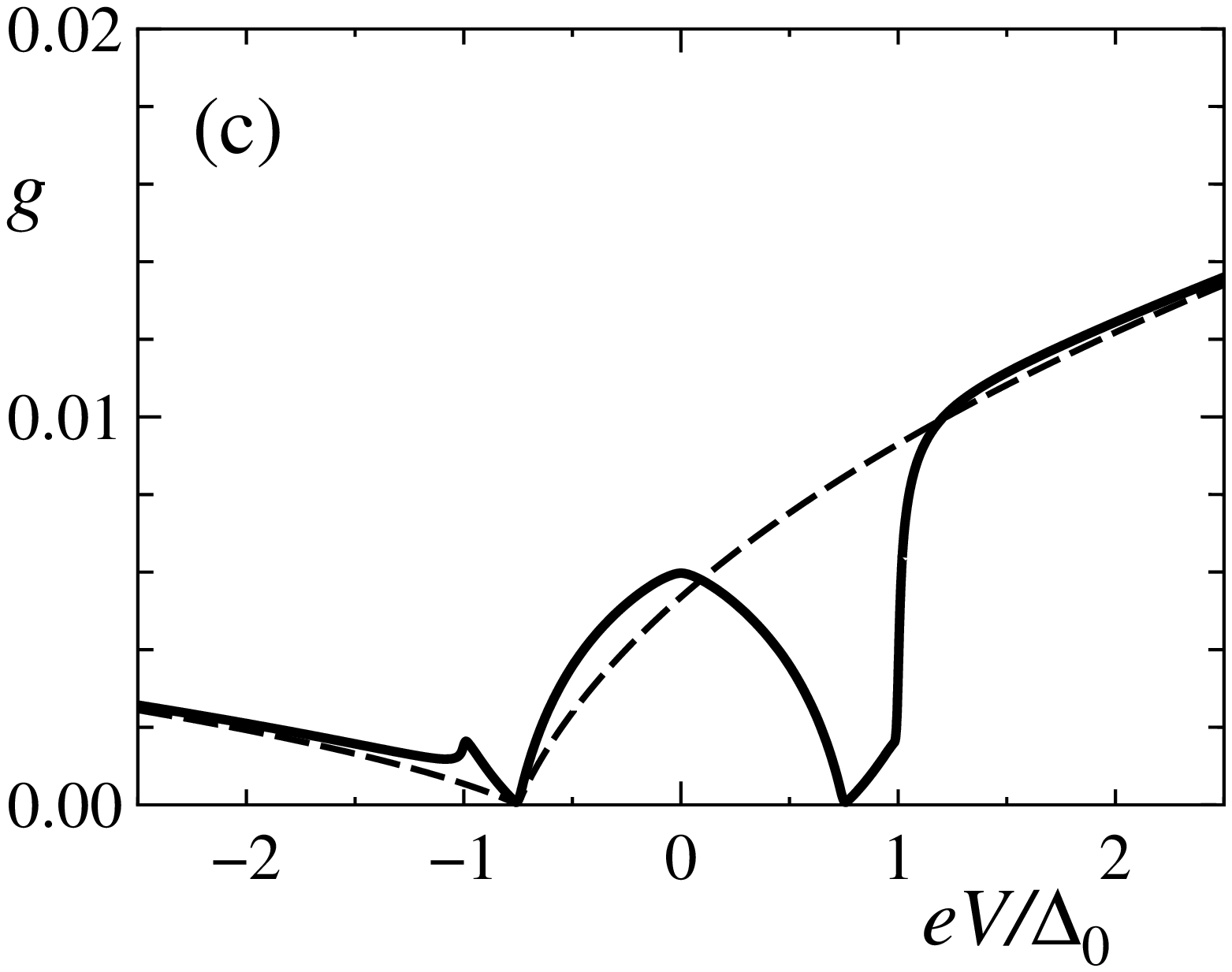}
\includegraphics[height=3.3cm]{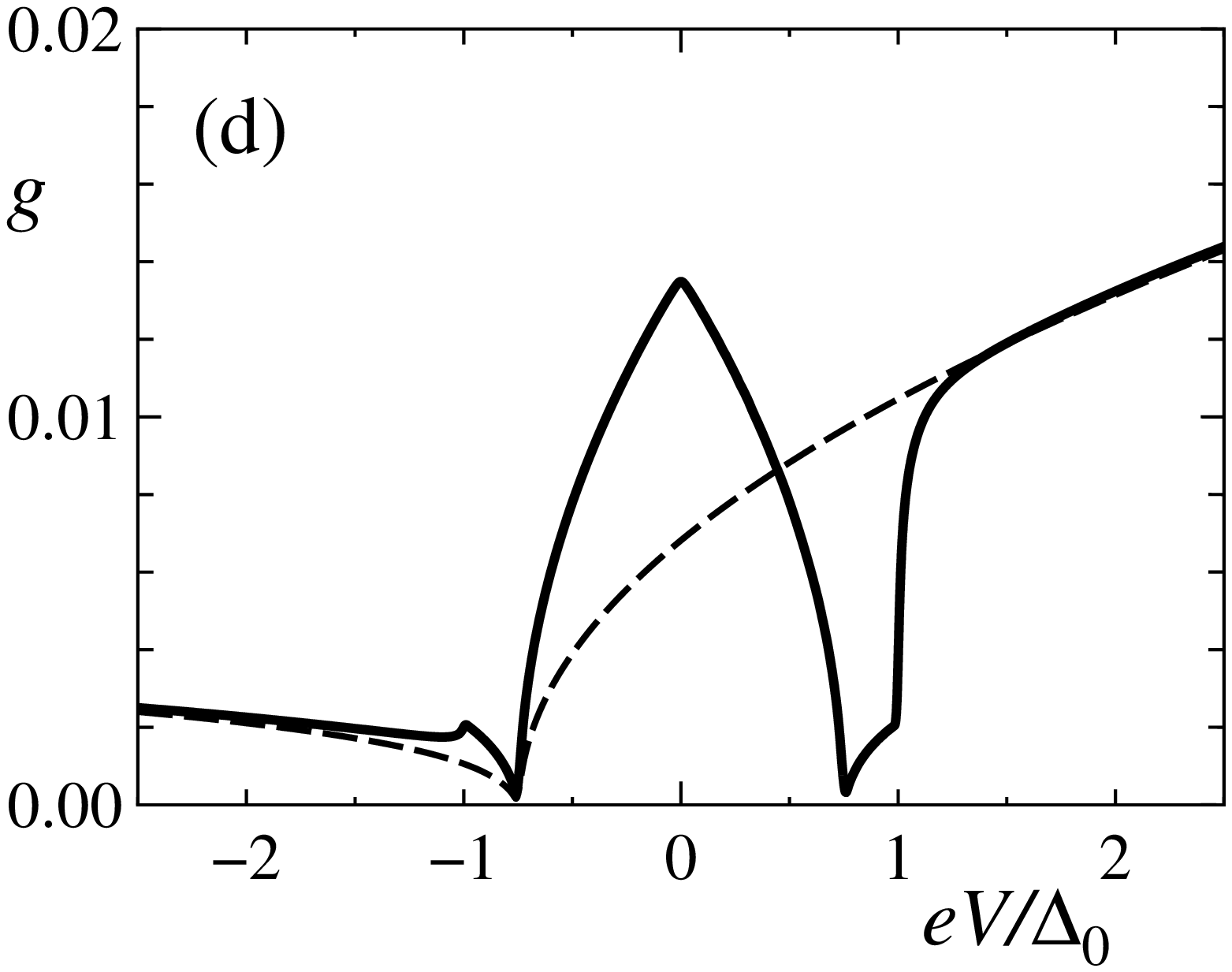}
\includegraphics[height=3.3cm]{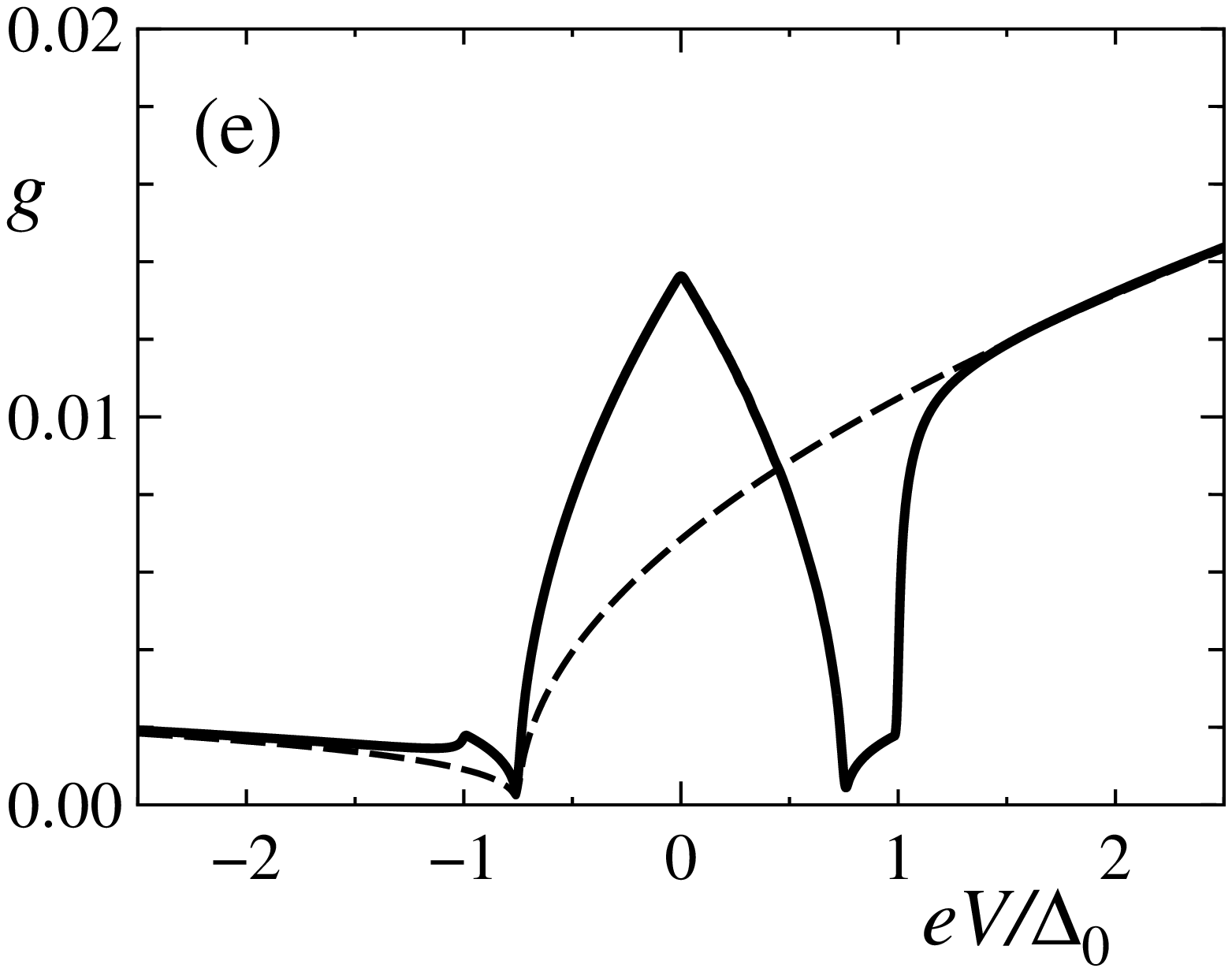}
\includegraphics[height=3.3cm]{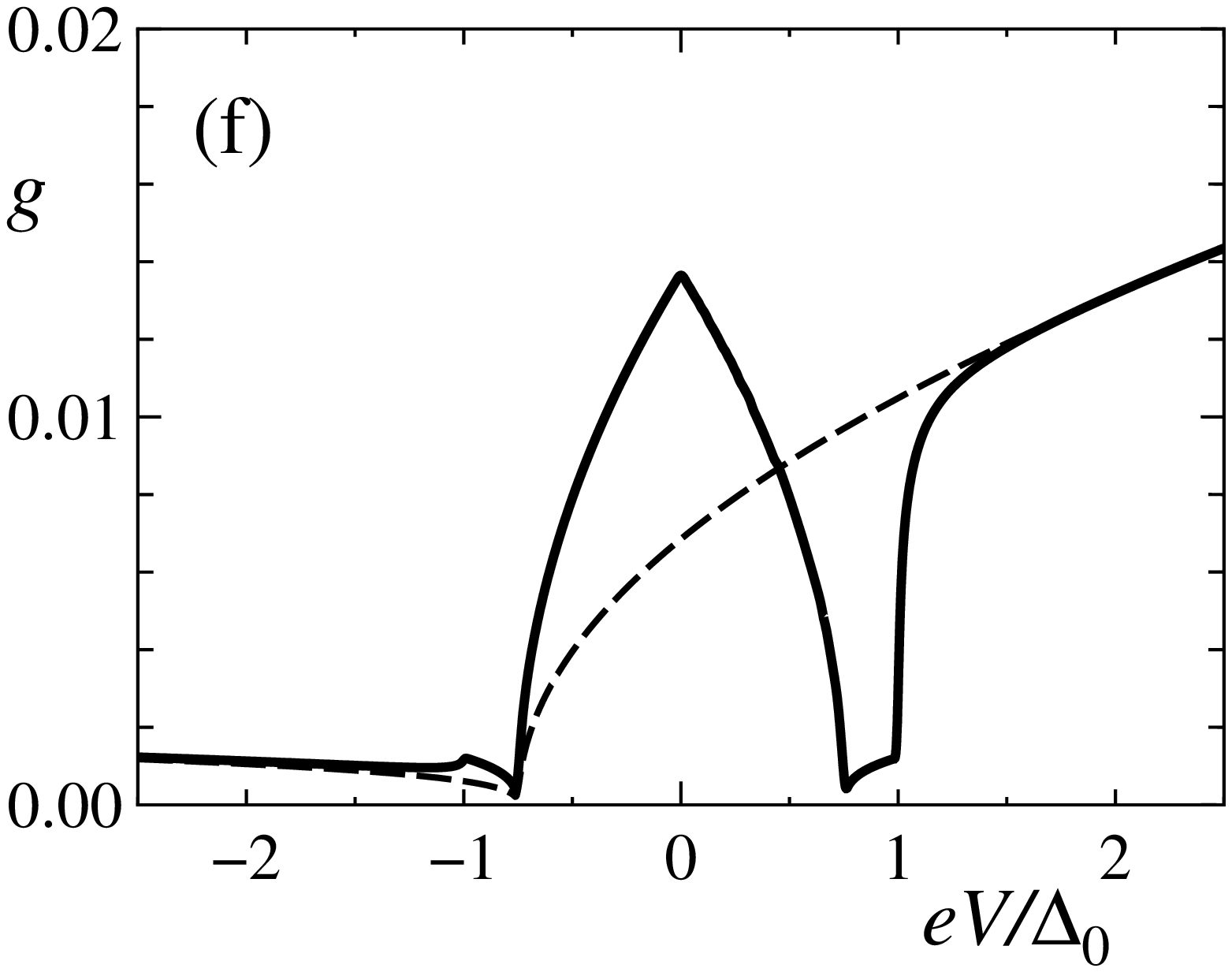}
\end{center}
\caption{
$eV$ dependence of $g_{\rm NS}(E_{\rm F}+eV)$
at $E_{\rm F}/\Delta_{0} = 0.75$ with (a) $\lambda = 0.0$, (b) $2.46$,
(c) $4.92$, (d) $40$, (e) $100$, and (f) $400$ nm.
}
\end{figure}

\begin{figure}[tb]
\begin{center}
\includegraphics[height=3.3cm]{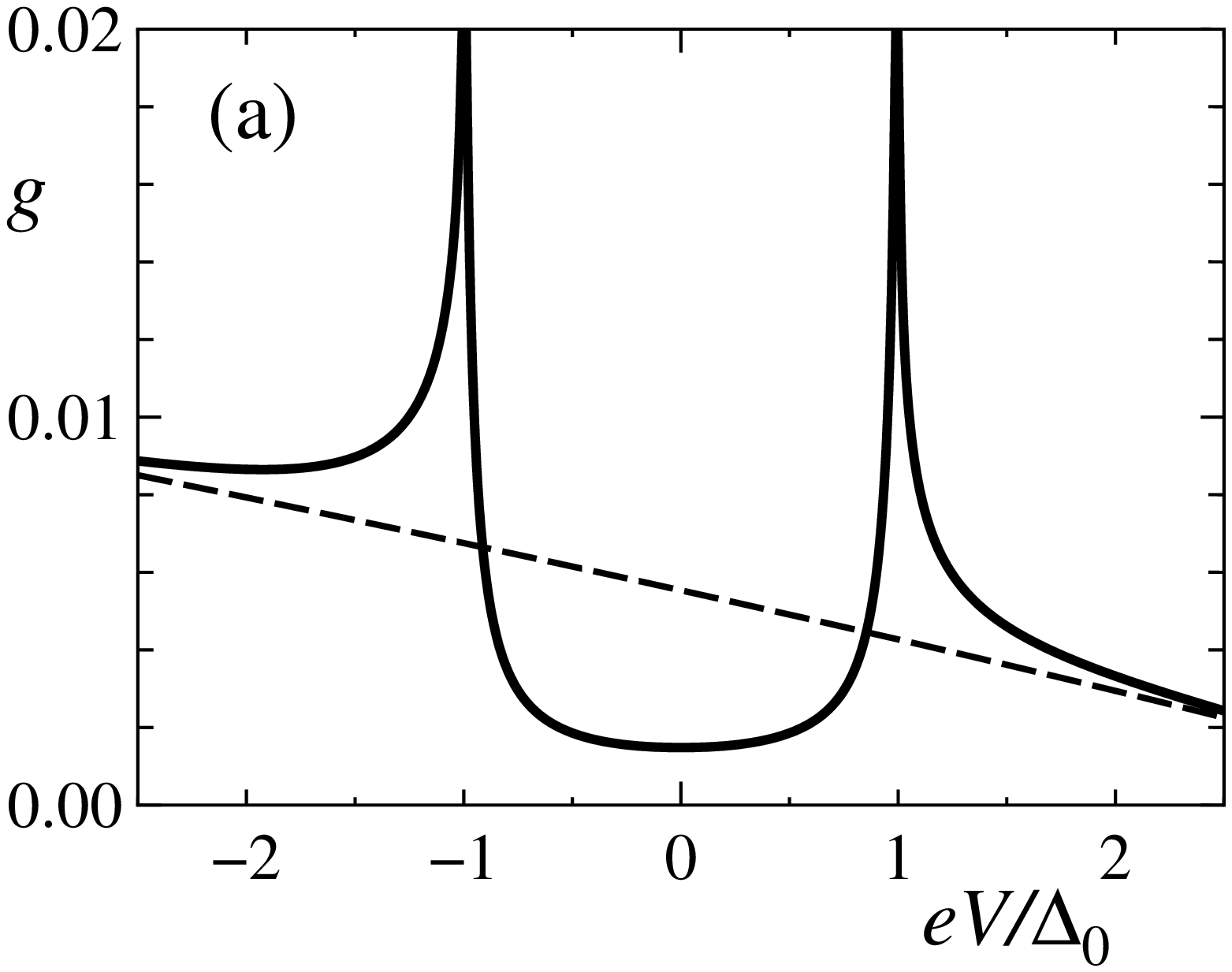}
\includegraphics[height=3.3cm]{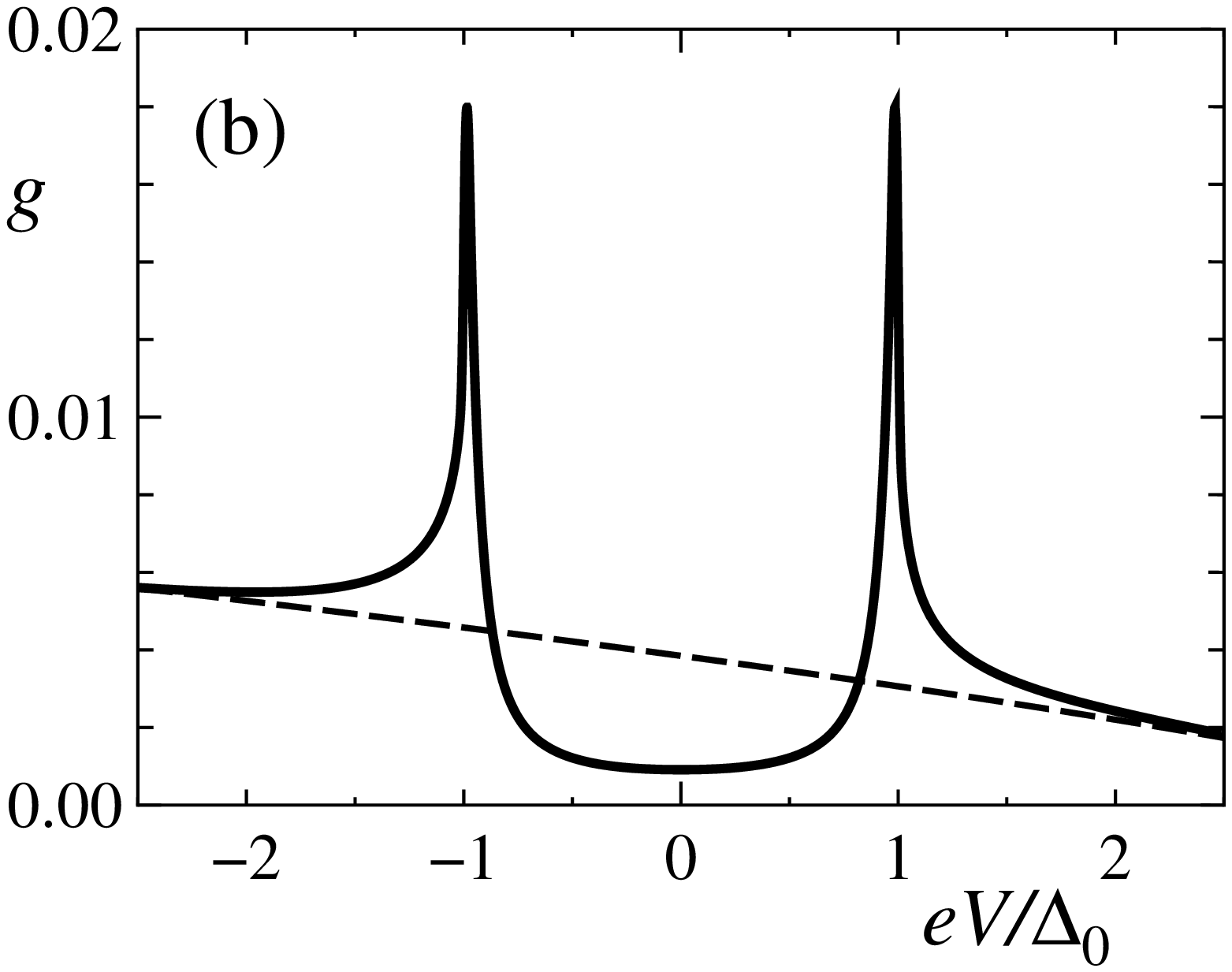}
\includegraphics[height=3.3cm]{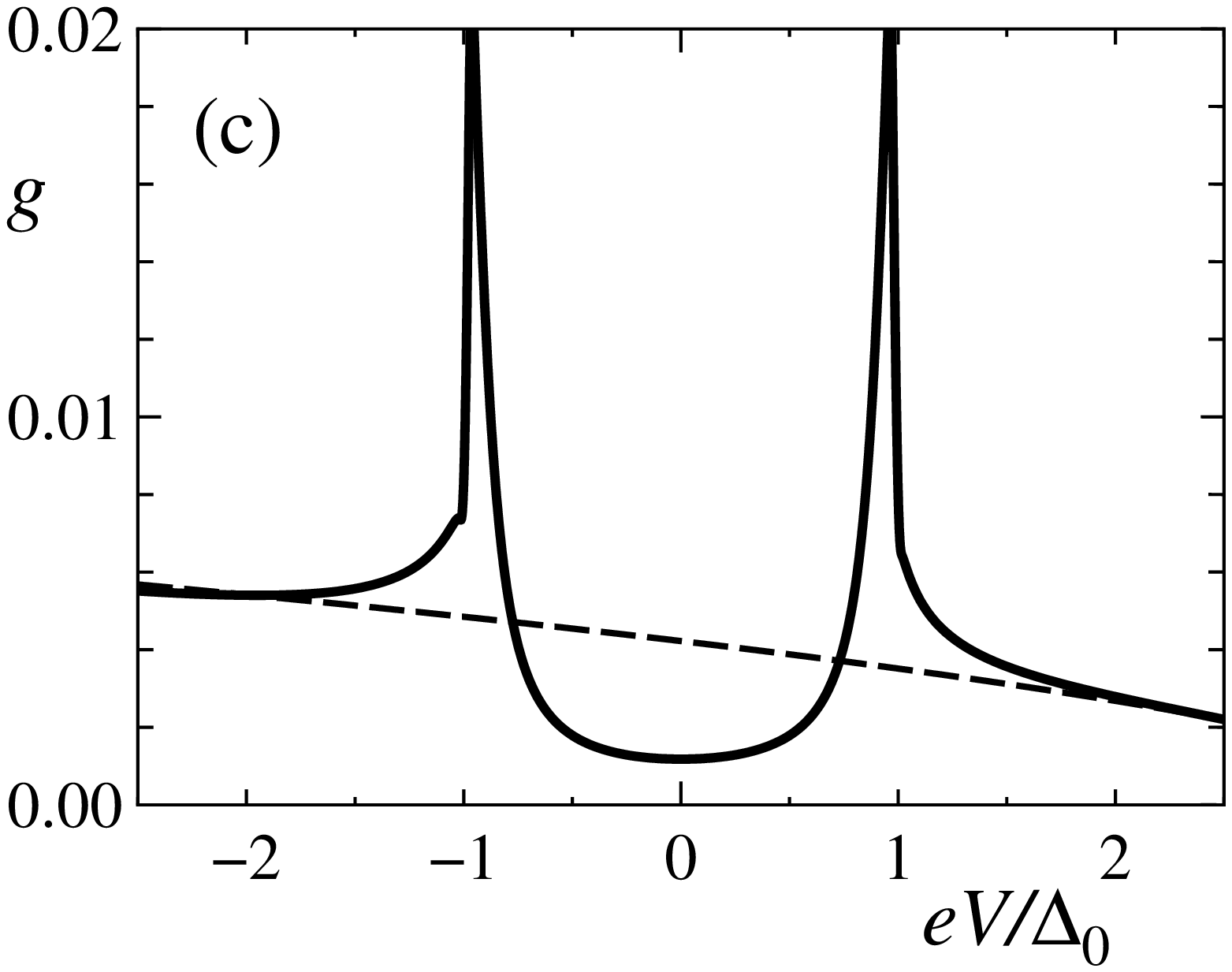}
\includegraphics[height=3.3cm]{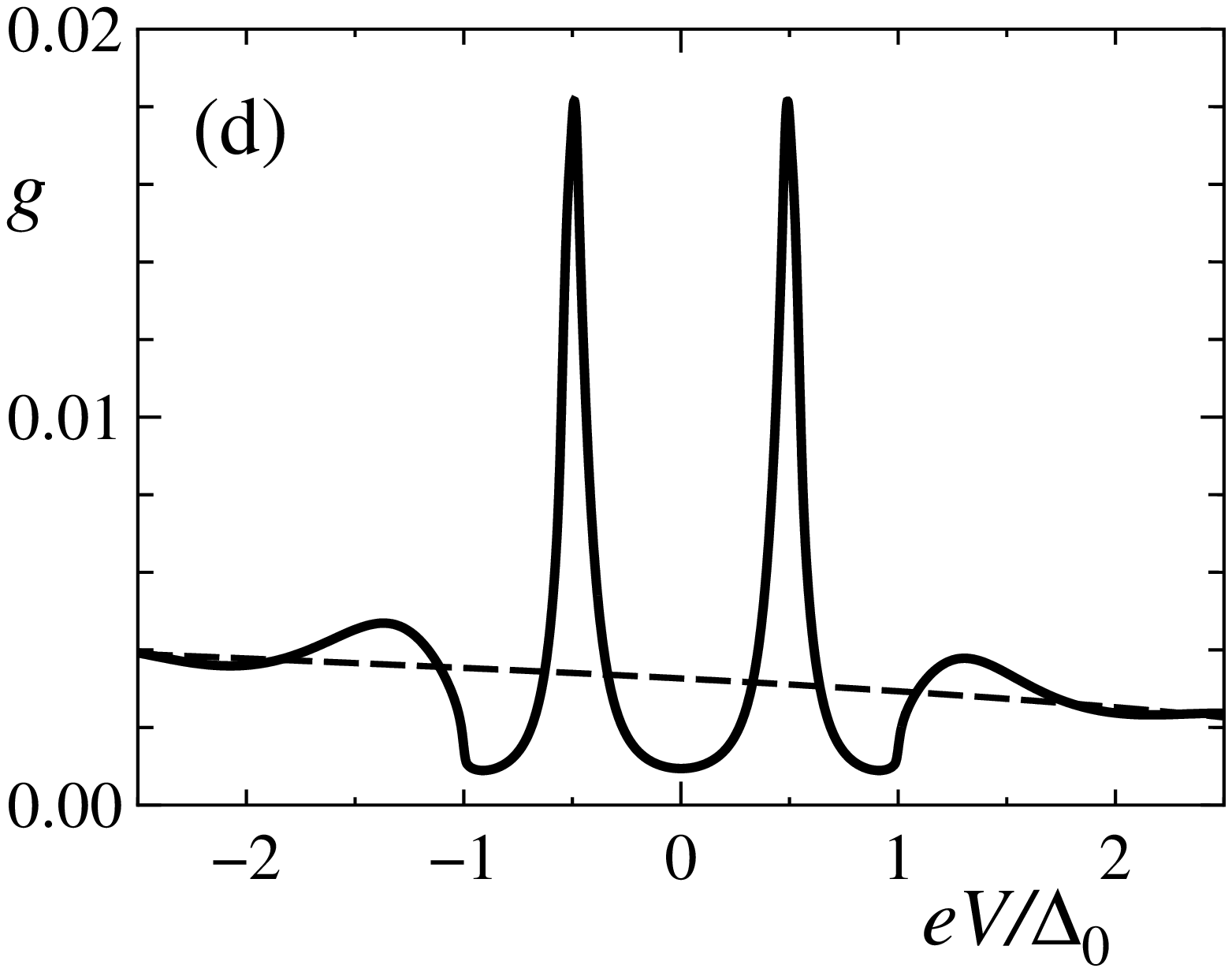}
\includegraphics[height=3.3cm]{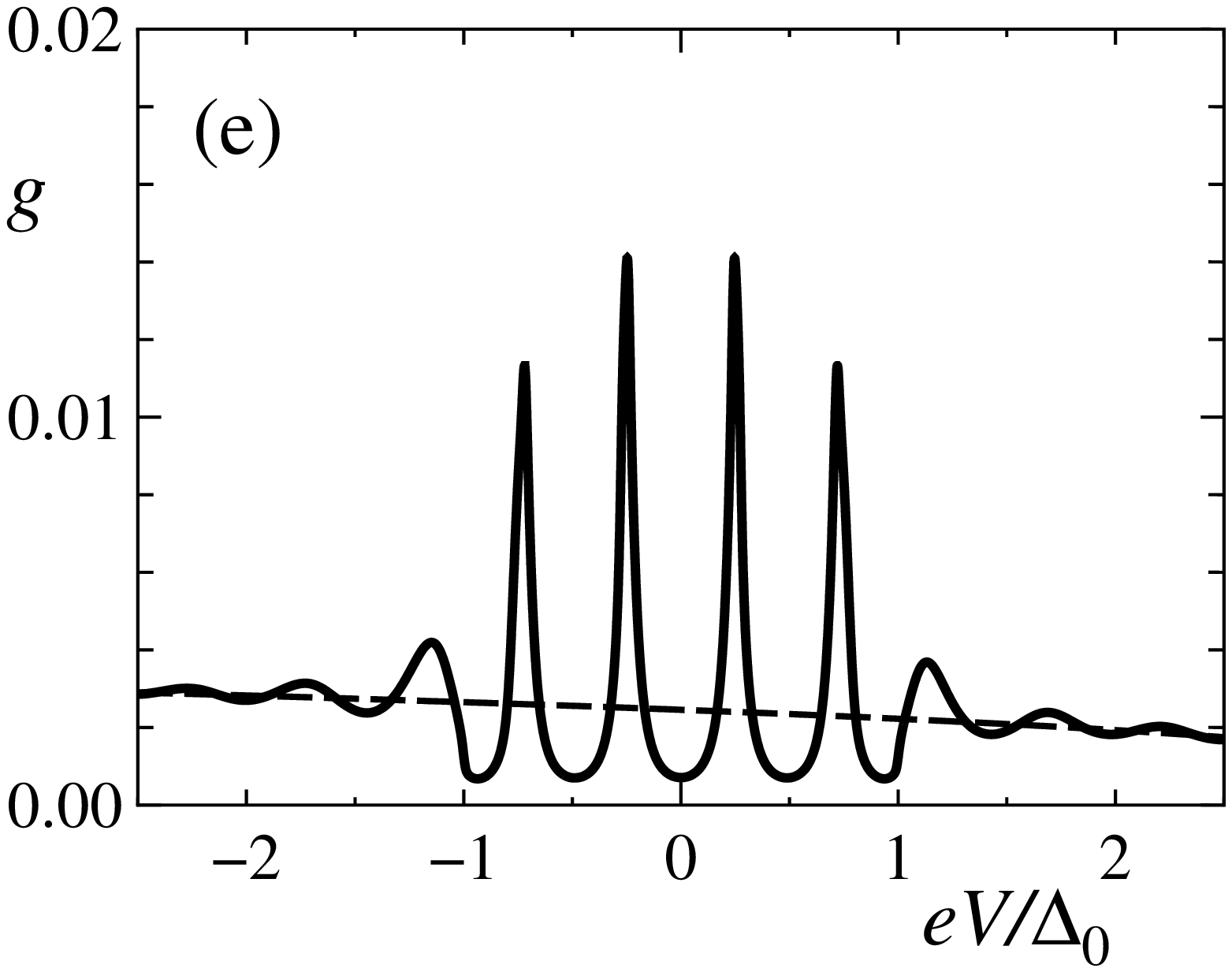}
\includegraphics[height=3.3cm]{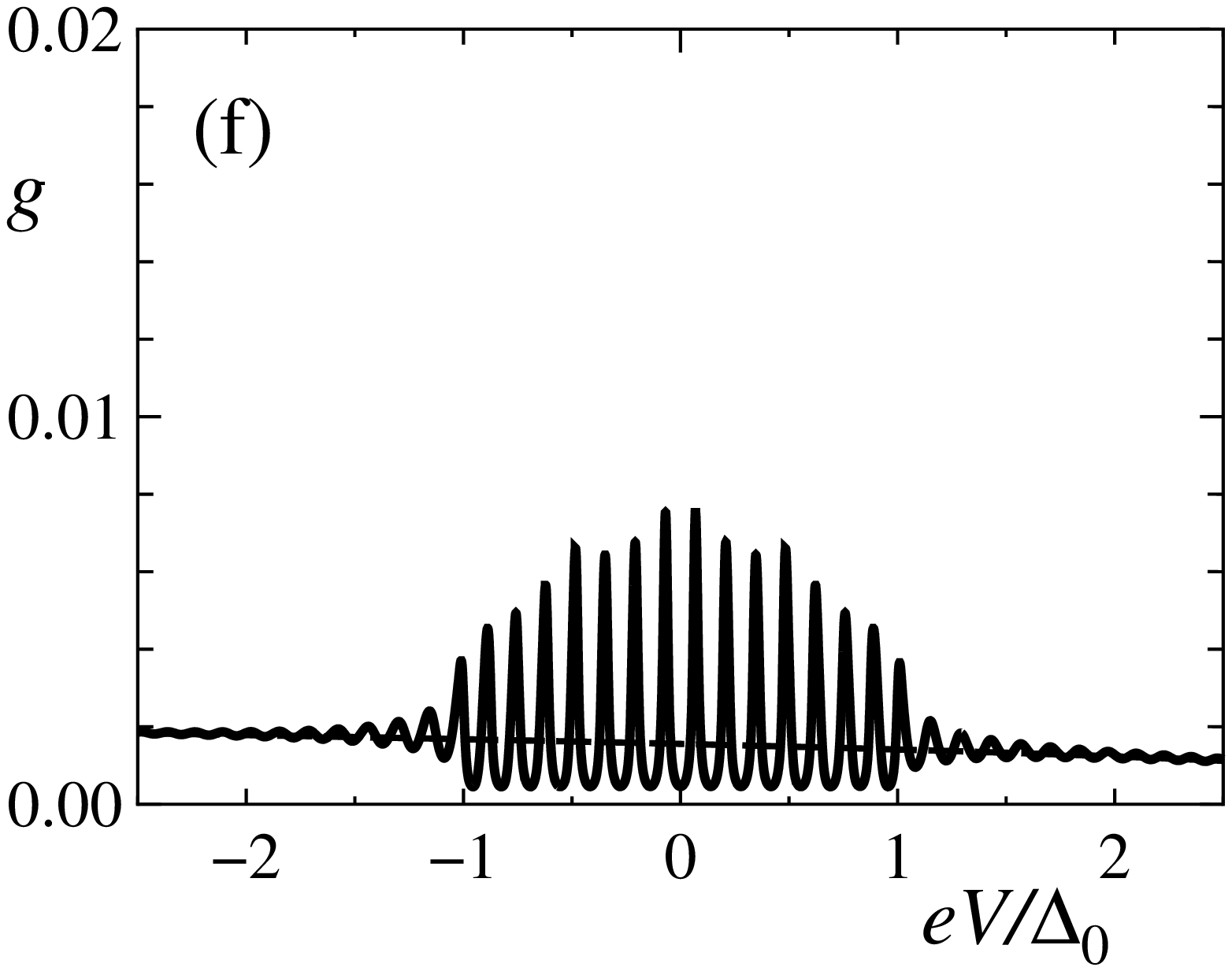}
\end{center}
\caption{
$eV$ dependence of $g_{\rm NS}(E_{\rm F}+eV)$
at $E_{\rm F}/\Delta_{0} = -4.0$ for (a) $\lambda = 0.0$, (b) $2.46$,
(c) $4.92$, (d) $40$, (e) $100$, and (f) $400$ nm.
}
\end{figure}

Finally, we examine how the results shown above depend on
the screening length $\lambda$.
Figures~10, 11, and 12 show $g_{\rm NS}(E_{\rm F}+eV)$
at $E_{\rm F}/\Delta = 5.0$, $0.75$, and $-4.0$, respectively,
for $\lambda = 0.0$, $2.46$, $4.92$, $40$, $100$, and $400$ nm.
Figures~10 and 11 demonstrate that the differential conductance sensitively
depends on $\lambda$ when $\lambda$ is sufficiently small.
However, it becomes insensitive to $\lambda$
when $\lambda$ is $40$ nm or larger.
This behavior does not apply to the case of $E_{\rm F}/\Delta = -4.0$,
where quasi-bound states play a role.
Figure~12 demonstrates that $g_{\rm NS}(E_{\rm F}+eV)$
sensitively depends on $\lambda$ even when $\lambda$ is sufficiently large.
In the case of $\lambda < 40$ nm,
$g_{\rm NS}$ shows two peaks at the gap edges,
implying that quasi-bound states are not created near the NS interface.
Contrastingly, $g_{\rm NS}$ shows resonant peaks in the case of
$\lambda \ge 40$ nm, implying the existence of quasi-bound states.
The spacing between two successive peaks is reduced with increasing $\lambda$,
which is consistent with Eq.~(\ref{eq:est-QBS}) given below.

In the remainder of this section, we roughly estimate the location
of the resonant peaks observed in Fig.~12.
We suppose that quasi-bound states are created between the NS interface
and the $pn$ junction separated by a distance $\Lambda$ as shown in Fig.~9(b).
Let us consider the case where an electron with energy $E_{\rm F}+E$
is incident to the NS interface from the asymptotic regime of $x \le 0$.
We expect that its AR probability is enhanced when $E_{\rm F}+E$
becomes nearly equal to the energy of the quasi-bound states.
Adapting the argument of Ref.~\citen{kulik} to our problem,
the energy of the quasi-bound states is approximately determined by
\begin{align}
    \label{eq:est-QBS}
   E_{n}
   = \frac{\sqrt{3}\pi}{4}\frac{a\gamma_{0}}{\lambda}
     \sqrt{\frac{U_{0}}{\gamma_{1}}}
     \frac{\sqrt{\Sigma}}{\arctan\sqrt{\Sigma^{-1}-1}}
     \left(n+\varphi(E_{n})\right) ,
\end{align}
where $n$ is an integer, $\varphi(E) = \arccos(E/\Delta_{0})/\pi$, and
\begin{align}
   \Sigma = -\frac{E_{\rm F}(1+\sin\phi^{2})}{U_{0}} .
\end{align}
Here, $\phi$ is the angle of incidence in the asymptotic region.
The derivation of Eq.~(\ref{eq:est-QBS}) is given in the Appendix.
By analyzing the angle-resolved transmission probability of an electron
through a $pn$ junction, we find that the transmission probability is maximized
when $\phi$ is nearly equal to $\pi/4$.
In the case of $E_{\rm F}/\Delta_{0} = -4.0$ with $\lambda = 100$ nm
and $\phi = \pi/4$, Eq.~(\ref{eq:est-QBS}) becomes
\begin{align}
  \frac{E_{n}}{\Delta_{0}}
  \approx 0.8 \left(n+\varphi(E_{n})\right) .
\end{align}
This gives $E_{0}/\Delta_{0} \approx 0.32$ and $E_{1}/\Delta_{0} \approx 0.91$,
which roughly explain the result shown in Fig.~12(e).

\section{Summary and Discussion}

We numerically calculated the differential conductance
in a proximity junction of bilayer graphene
made by partially covering a graphene sheet by a bulk superconductor.
We took account of the spatial variation of the charge neutrality point (CNP)
caused by the inhomogeneity of the carrier concentration
in the uncovered region.
It was shown that, when the Fermi level is located close to the CNP,
the differential conductance shows an unusual asymmetric behavior
as a function of bias voltage.
This was attributed to a $pn$ junction naturally formed
in the uncovered region.
It was also shown that, if the Fermi level is located below the CNP,
resonant peaks appear in the differential conductance,
reflecting the presence of quasi-bound states created by the $pn$ junction
together with the NS interface.
This implies that one can estimate $\lambda$ if such a peak structure is
experimentally observed.

It is natural to question whether the phenomena mentioned above
also appear in a monolayer graphene junction.
A plausible answer is yes since a $pn$ junction, which is essential for their
appearance, can also be formed in it.
However, as the transport property of a $pn$ junction is qualitatively
different between the monolayer and bilayer cases,~\cite{comment2}
it is unclear how the phenomena in the two cases
resemble or differ from each other.
The monolayer case deserves separate consideration.

Here, we point out differences between the theoretical analyses of this study
and Ref.~\citen{efetov2}.
The most striking difference is that the spatial variation of the CNP
in the uncovered region was explicitly taken into account in this study
while the CNP was assumed to vary in a stepwise manner across
the NS interface in Ref.~\citen{efetov2}.
Hence, the effects of a $pn$ junction were not incorporated
in the analysis of Ref.~\citen{efetov2}.
Although the result in Sect.~3 in the case of $\lambda = 0$ corresponds to
that of Ref.~\citen{efetov2}, they are not equivalent in the sense
that intervalley scattering was not incorporated in the effective mass theory
of Ref.~\citen{efetov2}.
Another difference concerns the parameter $U_{0}$,
which was set equal to $1.0$ eV (i.e., $U_{0} > \gamma_{1}$) in this study
in accordance with the difference
between $\Phi_{\rm NbSe_2}$ and $\Phi_{\rm BG}$
while the case of $U_{0} \ll \gamma_{1}$ was mainly considered
in Ref.~\citen{efetov2}.

The numerical results shown in Sect.~3 clearly demonstrate that
the differential conductance in a bilayer graphene junction strongly depends
on the screening length $\lambda$
over which the location of the CNP spatially varies.
This raises a question of whether an analysis based on the model
with $\lambda = 0$ correctly describes experimental results
for a graphene junction.

\section*{Acknowledgment}

This work was supported by JSPS KAKENHI Grant Numbers JP15K05130,
and JP16H00897 [Science of Atomic Layers (SATL)].

\appendix

\section{Derivation of Eq.~(\ref{eq:est-QBS})}

Let $k_{+}(x)$ and $k_{-}(x)$ respectively be the $x$ dependent
longitudinal wave number for an electron with energy $E_{\rm F}+E$
and that for a hole with energy $E_{\rm F}-E$,
where both $E_{\rm F}+E$ and $E_{\rm F}-E$ are assumed to be negative.
They satisfy
\begin{align}
    \left| E+E_{\rm F}-V(x) \right|
  & = \frac{\left(\frac{\sqrt{3}}{2}\gamma_{0}a\right)^{2}}{\gamma_{1}}
      \left(k_{+}(x)^{2}+q^{2}\right) ,
          \\
    \left| E-E_{\rm F}+V(x) \right|
  & = \frac{\left(\frac{\sqrt{3}}{2}\gamma_{0}a\right)^{2}}{\gamma_{1}}
      \left(k_{-}(x)^{2}+q^{2}\right)
\end{align}
within effective mass theory.
In terms of the angle of incidence $\phi$ defined in the asymptotic region,
$q^{2}$ is expressed as
\begin{align}
   q^{2}
   = \frac{\gamma_{1}}{\left(\frac{\sqrt{3}}{2}\gamma_{0}a\right)^{2}}
     \left(-E-E_{\rm F}\right)\sin^{2}\phi .
\end{align}
Using this, we approximately obtain $k_{+}(x)$ and $k_{-}(x)$
in the region near the NS interface as
\begin{align}
  k_{+}(x)
  & = k_{\rm F}(x)
      + \frac{2\gamma_{1}}{(\sqrt{3}\gamma_{0}a)^{2}}
        \frac{E\left(1+\sin^{2}\phi\right)}{k_{\rm F}(x)} ,
         \\
  k_{-}(x)
  & = k_{\rm F}(x)
      - \frac{2\gamma_{1}}{(\sqrt{3}\gamma_{0}a)^{2}}
        \frac{E\left(1-\sin^{2}\phi\right)}{k_{\rm F}(x)} ,
\end{align}
where
\begin{align}
       \label{eq:k_F-x}
    k_{\rm F}(x)
    = \frac{2\sqrt{\gamma_{1}}}{\sqrt{3}\gamma_{0}a}
      \sqrt{E_{\rm F}\left(1+\sin^{2}\phi\right)-V(x)} .
\end{align}
Equation~(\ref{eq:k_F-x}) indicates that the crossing point is determined by
\begin{align}
      \label{eq:x_pn}
    E_{\rm F}\left(1+\sin^{2}\phi\right)= V(x_{\rm pn}) .
\end{align}

Now, we consider the quantization of a quasiparticle confined in a region
of length $\Lambda \equiv L - x_{pn}$ near the NS interface
on the basis of the argument in Ref.~\citen{kulik}.
The quantization condition is expressed as
\begin{align}
    \xi^{2}\exp \left[2i\int_{L-\Lambda}^{L}dx\left(k_{+}(x)-k_{-}(x)\right)
                 \right] = 1
\end{align}
with
\begin{align}
  \xi = \frac{E-i\sqrt{\Delta_{0}^{2}-E^{2}}}{\Delta_{0}} .
\end{align}
After performing the integration over $x$, we find that
the energy of quasi-bound states $E_{n}$ satisfies
\begin{align}
   E_{n}
   = \frac{\sqrt{3}\pi}{4}\frac{a\gamma_{0}}{\lambda}
     \sqrt{\frac{U_{0}}{\gamma_{1}}}
     \frac{e^{-\frac{\Lambda}{2\lambda}}}
          {\arctan\sqrt{e^{\frac{\Lambda}{\lambda}}-1}}
     \left(n+\varphi(E_{n})\right) ,
\end{align}
where $\varphi(E) = \arccos(E/\Delta_{0})/\pi$.
The above equation is rewritten in the form of Eq.~(\ref{eq:est-QBS})
by using Eq.~(\ref{eq:x_pn}) rewritten in the form of
\begin{align}
  e^{-\frac{\Lambda}{\lambda}} = -\frac{E_{\rm F}(1+\sin^{2}\phi)}{U_{0}} .
\end{align}


\begin{thebibliography}{99}
\bibitem{comment1} The reduction of $\Delta(x)$ near the NS interface
is ignored in Eq.~(\ref{eq:Delta-def}).
It is not expected that this is essential in considering qualitative features
of the electron transport in the presence of spatial variation of the CNP.
\bibitem{comment2} In the monolayer case, the transmission probability
of an electron through a $pn$ junction is maximized at $\phi = 0$.
\end{thebibliography}
\end{document}